\documentclass[fleqn,usenatbib]{mnras}
\usepackage{newtxtext,newtxmath}
\usepackage[T1]{fontenc}
\DeclareRobustCommand{\VAN}[3]{#2}
\let\VANthebibliography\thebibliography
\def\thebibliography{\DeclareRobustCommand{\VAN}[3]{##3}\VANthebibliography}

\usepackage{graphicx}	
\usepackage{amsmath}	

\usepackage{amssymb}	
\usepackage{multirow}
\usepackage{indentfirst}
\usepackage{url}

\newcommand{\be}{\begin{equation}}
\newcommand{\ee}{\end{equation}}
\newcommand{\ba}{\begin{eqnarray}}
\newcommand{\ea}{\end{eqnarray}}
\newcommand{\bi}{\begin{itemize}}
\newcommand{\ei}{\end{itemize}}

\title{Detection of pairwise kSZ effect with DESI galaxy clusters and \emph{Planck}}

\author[Chen et al.]{
Ziyang Chen,$^{1,2}$\thanks{E-mail: chen\_zy@sjtu.edu.cn}
Pengjie Zhang,$^{1,2,3}$\thanks{E-mail: zhangpj@sjtu.edu.cn} 
Xiaohu Yang,$^{1,2,3}$
Yi Zheng,$^{4,2}$
\\
$^1$Department of Astronomy, School of Physics and Astronomy, Shanghai Jiao Tong University, Shanghai, 200240, China\\ 
$^2$Key Laboratory for Particle Astrophysics and Cosmology (MOE)/Shanghai Key Laboratory for Particle Physics and Cosmology, China\\
$^3$Tsung-Dao Lee Institute, Shanghai Jiao Tong University , Shanghai
200240, China\\
$^4$School of Physics and Astronomy, Sun Yat-sen University, 2 Daxue Road, Tangjia, Zhuhai, 519082, China
}
\date{Accepted XXX. Received YYY; in original form ZZZ}

\pubyear{2021}

\begin{document}
\label{firstpage}
\pagerange{\pageref{firstpage}--\pageref{lastpage}}
\maketitle

\begin{abstract}
We report a $5\sigma$ detection of the pairwise kinematic Sunyaev-Zel'dovich (kSZ) effect, combining galaxy clusters in DESI imaging surveys and the Planck temperature maps. The detection is facilitated by both improvements in the data and in the analysis method. For the data, we adopt the recently released DESI galaxy group catalog with $\sim 10^6$ robustly-identified groups, and construct various galaxy cluster samples for the kSZ measurement. The DESI group catalogue also provides estimation of halo mass, which further improves the kSZ measurement by $\sim 10\%$. For the analysis method, we derive an optimal estimator of pairwise kSZ through the maximum likelihood analysis. It also handles  potential systematic errors self-consistently.  The baseline cluster sample,  containing the $1.2\times 10^5$ richest galaxy clusters of typical mass $\sim 10^{14} M_{\sun}/h$ at typical redshift $0.2$-$0.5$, rules out the null hypothesis at $5\sigma$.  When fitting with a pairwise kSZ template from simulations,  the signal is detected at $4.7\sigma$ and  the average optical depth is constrained as $\bar{\tau}_e=(1.66\pm 0.35)\times 10^{-4}$.  We perform various internal checks, with different cluster selection criteria, different sky coverage and redshift range, different CMB maps, different  filter sizes, different treatments of potential systematics and the covariance matrix. The kSZ effect is consistently detected with $2.5\leq $S/N$\leq 5.6$ and acceptable $\chi^2_{\rm min}$,  across a variety of cluster samples.  The S/N is limited by both the Planck resolution and the photo-z accuracy, and therefore can be significant improved with DESI spectroscopic redshift information and with other CMB experiments. 
\end{abstract}

\section{Introduction} \label{sec:intro}
The kinematic Sunyaev-Zel'dovich (kSZ) effect \citep{SZ72,1980MNRAS.190..413S} is rich in cosmological information \citep{2002ARA&A..40..643C,2021A&A...653A.130K}. Firstly, it directly probes missing baryons at low redshift and the associated thermodynamics (e.g.  \citet{2016MNRAS.458.3773S,2021PhRvD.103f3514A}). Secondly, the contribution from the epoch of reionization to kSZ is comparable to the later time contribution (e.g. \citet{2006NewAR..50..909I,2007ApJ...660..933I,2013ApJ...776...83B,2016ApJ...824..118A}). Therefore kSZ is also a powerful probe of reionization (e.g. \citet{2004MNRAS.347.1224Z,2005ApJ...630..643M,2016MNRAS.463.2425M,2021MNRAS.500..232P}). Thirdly, the kSZ effect is proportional to the large scale peculiar velocity, and can therefore  constrain dark energy or modifications to general relativity \citep{2008MNRAS.388..884Z,2014PhRvD..89f3009P,2020ApJ...904...48Z,2020arXiv201003762W,2021MNRAS.501.4565M}. Furthermore, the kSZ effect is sensitive to the CMB dipole seen by distance electrons, making it a powerful probe of the Copernican principle, inflation and CMB anomalies \citep{2011PhRvL.107d1301Z,2014A&A...561A..97P,2015JCAP...06..046Z,2020PhRvD.101l3508C}\\

However, kSZ measurement is still challenging, due to the  degenerate spectrum with primary CMB, weak signal concentrated at small angular scales, limited frequency coverage and contaminations of thermal Sunyaev-Zel'dovich effect (tSZ) and cosmic infrared background.  For such reasons, the power spectrum of the diffuse kSZ effect has been measured only at $\la 3\sigma$ level, even with multi-frequency and high resolution CMB experiments such as ACT and SPT \citep{2013JCAP...07..025D,2015ApJ...799..177G,2021ApJ...908..199R}. Fortunately, the large scale distribution of galaxies and galaxy clusters in overlapping sky are correlated with the kSZ effect, and have facilitated the kSZ detection significantly. On one hand, the kSZ effect of individual clusters has been detected \citep{2013ApJ...778...52S, 2016ApJ...820..101S,2017A&A...598A.115A}.  On the other hand, statistical detections of kSZ with the aid of galaxies/clusters have been successful. By stacking many galaxies or galaxy clusters, pairwise kSZ effect has been detected statistically over the last decade, combining galaxy surveys such as BOSS and DES, and CMB surveys of ACT\citep{2012PhRvL.109d1101H,2017JCAP...03..008D,2021arXiv210108374C}, Planck\citep{2016A&A...586A.140P,2018MNRAS.475.3764S,2018PhRvD..97b3514L} and SPT \citep{2016MNRAS.461.3172S}. With extra peculiar velocity information reconstructed through 3D galaxy distribution, the kSZ measurement can be further improved \citep{2016PhRvD..93h2002S,2020ApJ...889...48L,2021A&A...645A.112T,2021PhRvD.103f3513S}. Other methods, also with the aid of galaxy/galaxy cluster information, have also been applied in the data analysis \citep{2016PhRvL.117e1301H,2018A&A...617A..48P,2021MNRAS.503.1798C}.

One key ingredient in improving the pairwise kSZ measurement is to construct a sufficiently large catalog of robustly identified galaxy clusters. The recently released  DESI DR8 galaxy group catalog (\citet{2021ApJ...909..143Y}, hereafter Y21) covers $40\%$ of the sky and $z<1$. It is nearly complete for massive clusters. Therefore it provides an excellent opportunity to improve the pairwise kSZ measurement. In this work, we measure the pairwise kSZ effect  using Planck CMB maps and galaxy clusters selected from Y21.  

This paper is organized as follows. \S \ref{sec:theory} presents the formulas of pairwise kSZ effect and the  new estimator to measure it. \S  \ref{sec:data} describes the data set and analysis methods. \S \ref{sec:result}  presents measurements results, along with various tests. \S \ref{sec:conclu} concludes with discussions on possible improvements in the near future. We include an appendix to explain further details and to perform more tests of the kSZ data analysis. 

\section{kSZ Theory} \label{sec:theory}
The change of CMB temperature caused by the kSZ effect is 	
\ba
	\label{eq:kSZ_general}
	\frac{\Delta T _{\rm kSZ}}{T_{\rm CMB}}(\hat{\textbf n})=-\int n_e(\hat{\textbf n},l)\sigma_T\frac{\hat{\textbf n}\cdot\textbf{v}}{c}dl
\ea	
\citep{1980MNRAS.190..413S}. Here, $n_e$ is the electron number density and $\hat{\textbf n}\cdot\textbf{v}$ is electron velocity projected along the line-of-sight direction. Assuming each CMB photon is only scattered by electron at most one time until arriving the observer, the kSZ effect generated by a galaxy group $i$ is  
\ba
	\label{eq:kSZ_a_group}
	\frac{\Delta T_{i,{\rm kSZ}}}{T_{\rm CMB}}(\hat{\textbf n}_i)=-\bar\tau \frac{\hat{\textbf n}_i\cdot\textbf v_i}{c}
\ea	
where $\tau$ is the mean optical depth of the group sample describing the baryon abundance associated with the group.
\subsection{Pairwise kSZ} \label{subsec:theory_pairwise_kSZ}
Due to the tendency that two clusters move towards each other under the influence of gravity, there is a net difference between $\Delta T_{\rm kSZ}$ of cluster pairs. Namely,  
\ba
	T_{\rm pkSZ}({\bf r})\equiv \langle \Delta
 	T_{i,{\rm kSZ}}({\bf x}+{\bf r})-\Delta
	T_{j,{\rm kSZ}}({\bf x})\rangle_{\bf  x}\neq 0\ . \nonumber
\ea
By symmetry, 
\ba
	\label{eq:kSZ_pairwise_2}
	T_{\rm pkSZ}({\bf r})&=&T_{\rm pkSZ}(r)\cos\theta \ .
\ea
$\theta$ is the angle between the pair separation and the line of sight.
Accurate modeling of $T_{\rm pkSZ}(r)$ requires accurate modeling of the cluster pairwise velocity $v_{12}$ and the optical depth-cluster bias relation. Nevertheless, given the $\sim 20\%$ accuracy in the pairwise kSZ measurement, we can conveniently adopt the following approximation (e.g. \citet{2018MNRAS.478.5320S}), 
\ba
	\label{eq:kSZ_pairwise_2_velocity_pairwise}
 	T_{\rm pkSZ}(r) &\simeq& -\bar{\tau} \frac{T_{\rm CMB}}{c}v_{12}(r)\ .
\ea
Here $\bar{\tau}$ is the average Thomson optical depth of clusters.  

The pairwise velocity $v_{12}(r)$ can be approximated with a theory curve based on linear perturbation (e.g. \citet{2010gfe..book.....M, Mueller_2015}) or adopt a template from N-body simulations. In this paper,  we use a pairwise velocity template from N-body simulation to fit the measured pairwise kSZ signal and obtain the mean optical depth. The simulation, one of  the ComicGrowth simulation \citep{Jing_2018},  has boxsize $L=1200\ {\rm Mpc}\cdot h^{-1}$ and  $3072^3$ particles. It adopts the WMAP cosmology, with $\Omega_{b}=0.0445$, $\Omega_{c}=0.2235$, $\Omega_{\Lambda}=0.732$, $h=0.71$, $n_s=0.986$, and $\sigma_8=0.83$. The halos are identified by Friends-to-Friends (FoF) algorithm, with a linking length b=0.2. The appendix \ref{app:simulation} shows the measured $v_{12}(r)$ of simulated halos in the mass and redshift range of observed clusters. 
	
\subsection{A new estimator of pairwise kSZ} \label{subsec:analysis_pairwise_kSZ_estimator}
A widely adopted estimator of pairwise kSZ effect, as adopted in the first detection by \citet{Hand_2012}, is 
\begin{eqnarray}
	\label{eq:pairwise_ksz}
	\hat{T}_{\rm pkSZ}(r)=-\frac{\sum_{i<j, r}T_{ij}c_{i j}}{\sum_{i<j, r} c_{i j}^{2}}\ . \nonumber
\end{eqnarray}	
Here $T_{ij}\equiv \Delta T_{i}-\Delta T_j$ is the temperature difference between the $i$-th and the $j$-th clusters. $c_{ij}\equiv \cos\theta_{ij}$ is the cosine between the average line of sight $(\vec{r_i}+\vec{r_j})/2$ and the pair separation $\vec{r}_{ij}\equiv \vec{r_i}-\vec{r}_j$. This estimator is motivated by Eq. \ref{eq:kSZ_pairwise_2_velocity_pairwise} and the weight $c_{ij}$ is to maximize the pairwise kSZ signal, which is $\propto c_{ij}$. However, it may be  biased in certain situations. Firstly in reality due to the limited case of finite cluster sample, $\langle c_{ij}\rangle\neq 0$.  For the same reason,  existing foregrounds in CMB (e.g. the cosmic infrared background (CIB)) have $\langle T^{\rm   foregrounds}_{ij}\rangle\neq 0$. This will bias the estimation of pairwise kSZ. Secondly, redshift dependent foregrounds such as CIB and tSZ and redshift dependent selections such as cluster mass/size cut and aperture filter may cause a redshift dependent $T_{ij}$. In such case, $\langle T_{ij}c_{ij}\rangle \neq 0$ and the kSZ measurement can be biased as well. Such potential bias is often corrected empirically by
\begin{eqnarray}
	\label{eq:model1}
	T_i\rightarrow T_{i}&-&
	\frac{\sum_{j} T_{j}G(z_{i}, z_{j}, \Sigma_{z})}{\sum_{j} G(z_{i}, z_{j}, \Sigma_{z})} \ . \nonumber
\end{eqnarray}
The function $G$ weighs over cluster temperatures around the given $i$-th cluster. A Gaussian form is often adopted, $G(z_{i}, z_{j}, \Sigma_{z})=\exp(-z_{ij}^2/2\Sigma_z^2)$ with $\Sigma_z\sim 0.01$. Here $z_{ij}\equiv z_i-z_j$. 

Given the above uncertainties in the conventional estimator, we decide
to construct a new estimator of the pairwise kSZ effect. It is required to maximize 
\begin{eqnarray}
	&&\mathcal{L} \propto \exp\left[-\frac{1}{2}\sum_{ij} \frac{(T_{ij}-T_{ij}^{\rm theory})^2}{\sigma^2_{ij}}\right]\ . 
\end{eqnarray}
Here $\sigma_{ij}$ is the r.m.s. error in $T_{ij}$ measurement.  $\sigma^2_{ij}=\sigma^2_i+\sigma^2_j$ and $\sigma_i$ is the noise of the temperature measurement of the $i$-th cluster.  For the theory, we include two kinds of contaminations. One is redshift independent ($n_0$) and the other depends on the pair redshift separation ($n_1$). 
\begin{eqnarray}\label{eq:model3}
	T_{ij}^{\rm theory}&=& \hat T_{\rm pkSZ}C_{ij}+\hat n_0+\hat n_1z_{ij}\ . 
\end{eqnarray}
Maximizing $\mathcal{L}$, we obtain the unbiased optimal estimator of $T_{\rm pkSZ}$, and also that of $n_{0,1}$,
\ba\label{eq:baseline_model}
	\left(
	\begin{array}{ccc}
		\langle C^2\rangle  & \langle CZ\rangle  & \langle C\rangle  \\
		\langle CZ\rangle   & \langle Z^2\rangle & \langle Z\rangle  \\
		\langle C\rangle    & \langle Z\rangle  & 1
	\end{array}
	\right)
	\left(
	\begin{array}{c}
		\hat T_{\rm pkSZ} \\ \hat n_1 \\ \hat n_0
	\end{array}
	\right)=
	\left(
	\begin{array}{c}
			\langle TC\rangle  \\\langle TZ\rangle \\\langle T\rangle 
	\end{array}
	\right)
\ea
For brevity, we have denoted $C_{ij}$ as $C$ and $z_{ij}$ as $Z$. The average $\langle ...\rangle $ is defined by
\ba
	\langle A\rangle \ \equiv\  \frac{\sum_{ij} A_{ij}/\sigma^2_{ij}}{\sum1/\sigma^2_{ij}}\ .
\ea

\subsection{Mass weight} \label{subsec:analysis_mass_weight}
The Y21 group catalog provides a good estimation of halo mass, with $\sim 0.2$-dex uncertainty. We know that a more massive halo trends to contain more baryons and generate stronger kSZ signal. We then expect
\ba
	\Delta T_{\rm kSZ} \propto M^{\alpha}\ .
\ea
If the gas fraction of all clusters are the same, $\alpha=1$. With such theory input, 
\begin{eqnarray}\label{eq:massw}
	T^{\rm theory}_{ij}&=&\tilde{T}_{\rm pkSZ}\tilde {c}_{ij}+n_0+n_1z_{ij}\ , \nonumber\\	
	\tilde {c}_{ij}&=&\frac{M^{\alpha}_i+M^{\alpha}_j}{2}c_{ij} \ .
\end{eqnarray}
The solution to $\tilde T_{\rm pkSZ}$ is identical to Eq.\ref{eq:baseline_model}, other than the replacements $\hat T_{\rm pkSZ}\rightarrow \tilde T_{\rm pkSZ}$ and $c_{ij}\rightarrow \tilde c_{ij}$. Once we obtain $\tilde T_{\rm pkSZ}$, we recover $\hat T_{\rm pkSZ}$ by
\ba
	\hat T_{\rm pkSZ}=\left\langle \frac{M^{\alpha}_i+M^{\alpha}_j}{2}\right\rangle \tilde{T}_{\rm pkSZ}\ .
\ea	

\section{Data and data analysis} \label{sec:data}
\subsection{Planck} \label{subsec:data_planck}
In this paper, we use a full-sky intensity map, HFI 217 GHz, of the public \textit {Planck} Release 3 data.\footnote{Based on observations obtained with Planck (http://www.esa.int/Planck), an ESA science mission with instruments and contributions directly funded by ESA Member States, NASA, and Canada.} The effective FWHM is 4.87 arcmin. This map is provided in HEALPix grid frame (\cite{2005ApJ...622..759G}) with $N_{\rm side}=2048$. Choosing 217 GHz is to minimize the tSZ contamination which vanishes at 217 GHz. Another advantage of this frequency band is the higher angular resolution than the foreground-cleaned CMB maps such as SMICA. This higher angular resolution is desired for the kSZ detection. This single frequency map may contain foregrounds other than primary CMB and kSZ. And the residual tSZ in 217GHz map is none zero due to the finite bandwidth. However, none of them is expected to have the directional dependence of pairwise kSZ. The designed estimator (Eq. \ref{eq:baseline_model}) can safely filter away such foregrounds and the measured pairwise kSZ will be free of such contaminations.  We also make measurements using the CMB foreground cleaned maps (SMICA, SEVEM, NILC, COMMANDER) in the Appendix \ref{app:different_cmb_maps}. We find that the 217GHz map is optimal for the kSZ measurement. 
\subsection{DESI DR8 galaxy groups/clusters} \label{subsec:data_desi}
\begin{figure}
 	\includegraphics[width=0.5\textwidth]{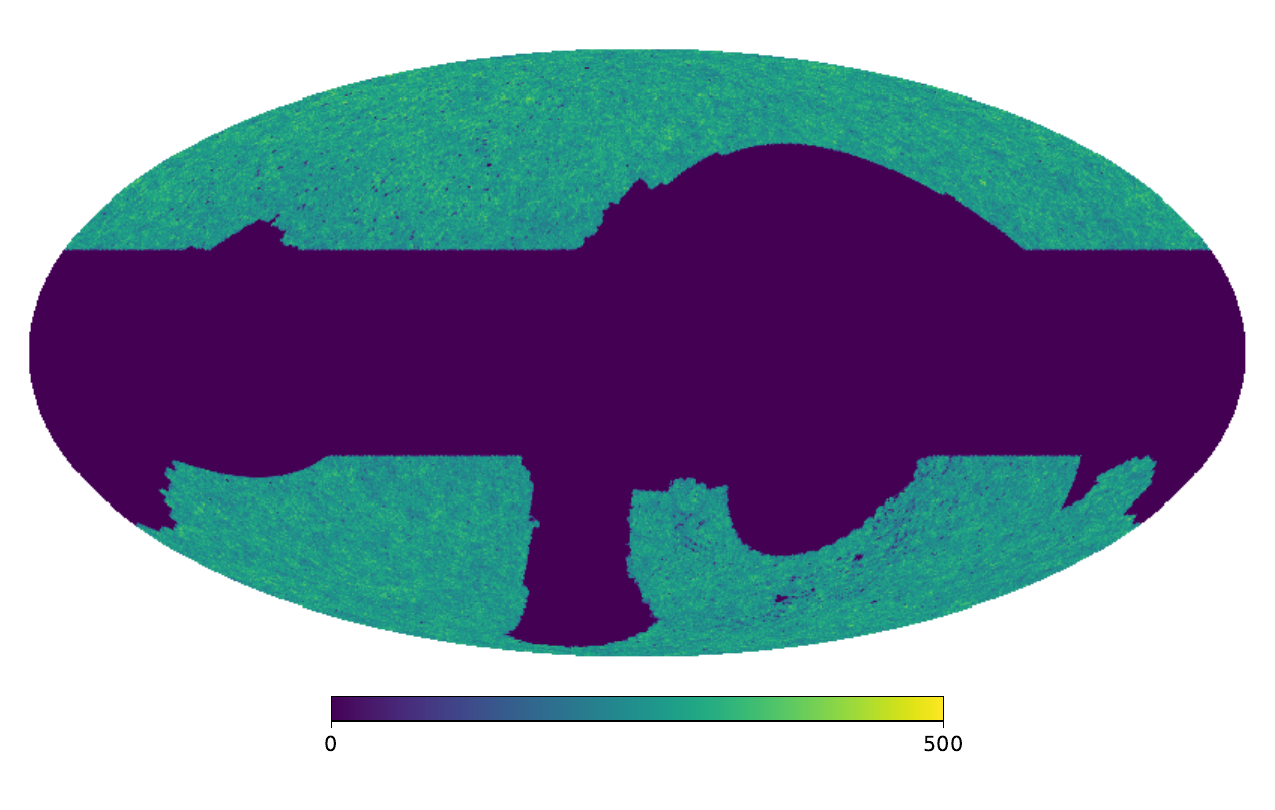}
	\caption{The group distribution of the DESI DR8 group catalog. The colorbar represents the number of groups included in Y21 in each pixel ($N_{\rm side}=256$). \label{fig:data_group_distribution}}
\end{figure}
We use a galaxy group catalog provided by \citet{2021ApJ...909..143Y}. This catalog is constructed from the Data Release 8 (DR8) of DESI Legacy Imaging survey, with an extended halo-based group finder developed by \citet{2005MNRAS.356.1293Y}. The sky area within $|b|<25\degr$ and around nearby sources or masked pixels have been removed before finding groups. The catalog contains 50.03 million galaxy groups in North galactic cap (NGC) and 42.26 million in the south galactic cap (SGC). The sky coverage is 9673 $\rm deg^2$ in NGC and 8580 $\rm deg^2$ in SGC (Fig. \ref{fig:data_group_distribution}). The catalog contains 3D coordinates, richness, halo masses and the total group luminosities. \\
\begin{figure}
	\includegraphics[width=0.45\textwidth]{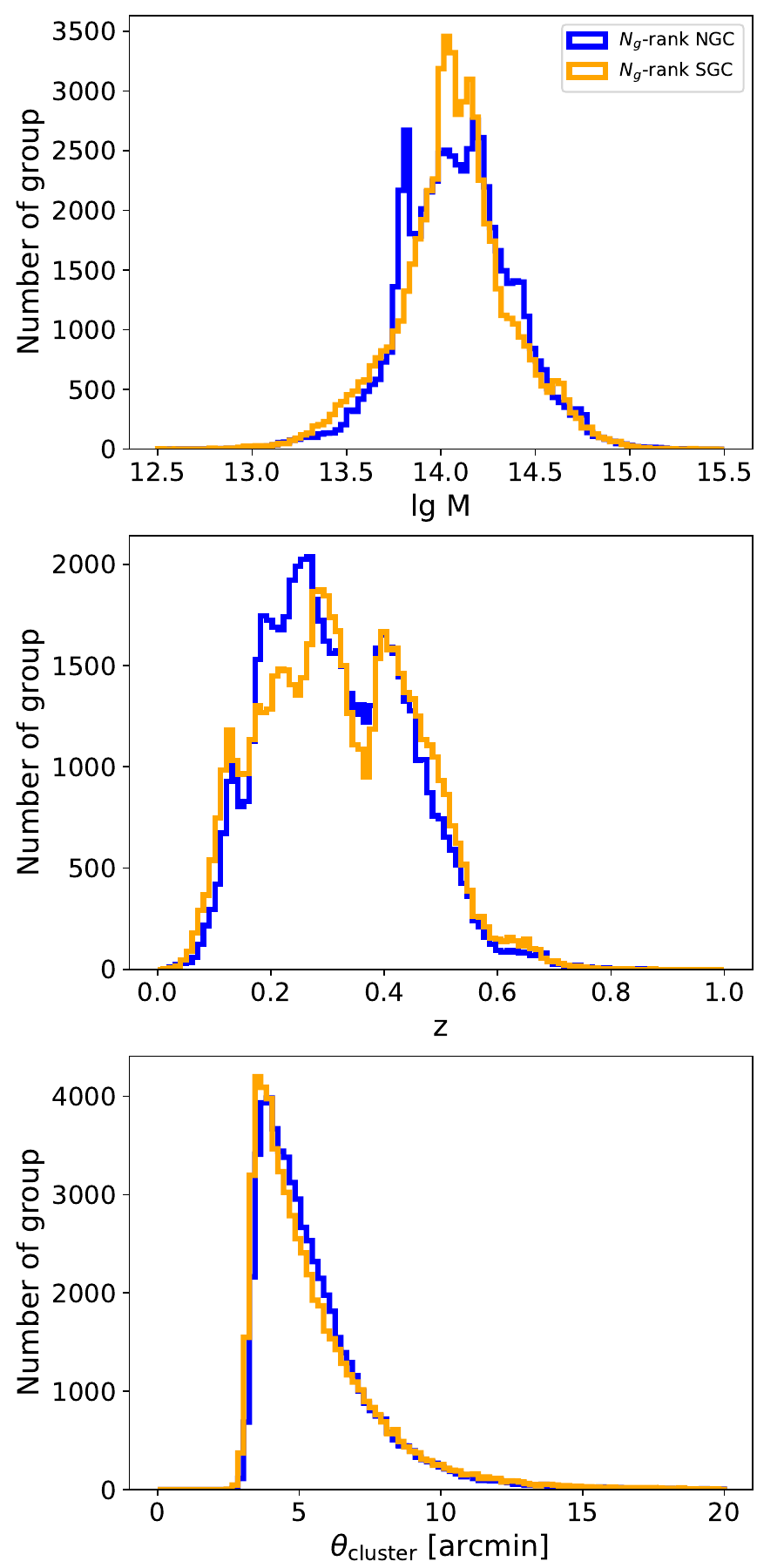}
	\caption{The mass, redshift and angular radius distribution of clusters in the baseline sample. Blue lines represent galaxy clusters in Northern Sky ($\bar z=0.32,\ {\rm lg}\bar M=14.19,\ \bar \theta=5.57\ [\rm arcmin]$), and yellow lines show those in Southern Sky  ($\bar z=0.32,\ {\rm lg}\bar M=14.17,\ \bar \theta=5.54\ [\rm arcmin]$). \label{fig:data_sample}}
\end{figure}	
We then construct galaxy cluster samples for the kSZ detection. There are four major considerations. (1) One is the richness $N_{\rm mem}$, namely the total number of member galaxies.  The purity of group catalogue increases with $N_{\rm mem}$. Furthermore, the gas  mass and therefore the kSZ effect are expected to increase with $N_{\rm mem}$. So this will be our primary selection criteria. The baseline cluster sample is composed of the $120,000$ richest galaxy clusters. This sample has a minimum $N_{\rm mem}=21$. Fig. \ref{fig:data_sample} shows the distribution of mass, redshift and angular radius within this sample. It also shows that halo mass peaks at $M\sim 10^{14}M_{\sun}/h$. The redshift range is almost between 0.1 and 0.6. And the minimum value of angular radius is 2.5 arcmin. So the baseline sample is largely composed of galaxy clusters with significant $\tau_e$. Indeed, this baseline sample enables kSZ detection at $5\sigma$. We have checked that, if $N_{\rm mem}$ is the only selection criterion, the baseline sample returns the highest S/N. (2) The second consideration is the halo mass $M_h$. Y21 provides estimation of halo mass by matching the luminosity distribution with the halo mass distribution. The total number of electrons and therefore the kSZ effect are expected to increase with increasing $M_h$, and sufficiently large halos, the relation is linear.  We will also use this extra information to improve the kSZ detection. However, since the estimated mass has  $\sim 0.2$-dex uncertainty, we do not include this criteria to define the baseline sample. (3) Cluster angular radius $\theta_{\rm cluster}$ also plays a role in kSZ detection. The beam of Planck temperature map has an effective FWHM $4.87^{'}$ at $217$ GHz. It is larger than $\theta_{\rm cluster}$ of a large fraction of groups/clusters, resulting into dilution of kSZ signal. Therefore larger $\theta_{\rm cluster}$, estimated by $M_h$, may lead to more significant kSZ detection. However, larger $\theta_{\rm cluster}$ could also mean lower redshift, lower mass and intrinsically smaller kSZ.  So we explore this criteria, but we do not use it to define our baseline sample. (4) Redshift also affects the kSZ detection, through the impacts in the purity of cluster sample, the accuracy of mass estimation, kSZ-richness relation, $\theta_{\rm cluster}$ and the dilution of kSZ. Due to the limited S/N of kSZ detection, we have to include clusters at all available redshifts. We only consider splitting clusters into two redshift bins for the purpose of checking the kSZ redshift dependence (Appendix \ref{app:redshift}). We also have a cluster sample with spectroscopic redshifts. 
	
In the main part of this paper, we will focus on the baseline cluster sample. We will also briefly show the cluster sample with highest S/N, and the cluster sample with spectroscopic redshifts. Results of other cluster samples, along with other options in the analysis, are presented in the Appendix \ref{app:choices_of_galaxy_sample}.	

\subsection{AP filter} \label{subsec:analysis_ap_filter}
To suppress the large scale noise, such as primary CMB, we apply an aperture photometry (AP) filter on the Planck CMB map.  
\begin{eqnarray}
	W_{\rm AP}(\theta)=\frac{1}{\pi\theta^2_{\rm AP}}
	\left\{
	\begin{array}{cc}
		1,&\quad \theta\leq\theta_{\rm AP}  \\
		-1,&\quad \theta_{\rm AP} < \theta \leq \sqrt{2}\theta_{\rm AP}\\
		0,&\quad \theta > \sqrt{2}\theta_{\rm AP} 
	\end{array}
	\right.
\end{eqnarray}
In this paper, we apply the AP filter in the spherical harmonic space, 
\begin{eqnarray}
	W_{\rm AP}(l)=\frac{2}{l\theta_{\rm AP}}[2J_1(l\theta_{\rm AP})-\sqrt{2}J_1(\sqrt{2}l\theta_{\rm AP})]\ . \nonumber\\
\end{eqnarray}
It peaks at $\ell\sim \pi/\theta_{\rm AP}$ (or $9000^{'}/\theta_{\rm AP}$ to be more accurate), and drops towards zero on both smaller or larger scales. We adopt a fiducial $\theta_{\rm AP}=3^{'}$, but we will also explore other choices \citep{2016PhRvD..94d3522A, 2018PhRvD..97b3514L}. 
	

\subsection{Photo-z
          correction} \label{subsec:theory_photo_z_corr}
The pairwise method need precise relative positions of a pair of clusters. Unfortunately redshifts of Y21 group catalog are mostly photometric-redshifts, with errors $\sigma_z/(1+z)\simeq 0.01$ \citep{yang2020extended}.  Photo-z error leads to errors/smoothing in $r$ or even alter the order of a pair of cluster in redshift space.  This would significantly suppresses the kSZ signal at $r\lesssim \sqrt{2}\sigma_z c/H\sim 30h^{-1}$ Mpc. This suppression is often modeled empirically. Here we include the effect of photo-z in our template through simulations. 	 In Appendix \ref{app:simulation}, we choose a halo sample whose redshift and bias are close to that of our baseline sample. Then we add a random shifting to halo positions along line-of-sight direction and calculate the pairwise velocity of them as the fitting template.



\subsection{The measurement S/N} \label{analysis_covariance_matrix}
We fit the measured $T_{\rm pkSZ}$ against the theoretical template. The $\chi^2$ is
\begin{eqnarray}
		\chi^{2}(\bar{\tau}_{e})
		&=&[\hat{T}_{\rm pkSZ}-\bar\tau _e \frac{T_{\rm CMB}}{c} v_{12}]^{\dagger} \hat{C}^{-1} \nonumber \\
		&\times& [\hat{T}_{\rm pkSZ}-\bar\tau _e
                   \frac{T_{\rm CMB}}{c} v_{12}]\ .
\end{eqnarray}
Notice that the term in the bracket is the data vector of size $N_{\rm r-bins}$. $C$ is the covariance matrix, estimated using jackknife resampling.  The minimum $\chi^2$ corresonds  to the best fit of $\bar{\tau}_e$, the mean optical depth of our cluster sample. Since it is a linear fitting with a single parameter $\bar{\tau}_e$, both the best-fit value and the associated statistical error are given analytically, 
\ba 
	\bar\tau _e^{\rm bestfit}=\frac{\hat{T}_{\rm pkSZ}^{\dagger}\hat{C}^{-1} T^{\rm theory}_{\rm pkSZ}}{{T}^{\rm theory \dagger}_{\rm pkSZ}\hat{C}^{-1}T^{\rm theory}_{\rm pkSZ}} \ ,
	\nonumber \\	
	\sigma^2_{\bar\tau_e}=\frac{1}{{T}^{\rm theory \dagger}_{\rm pkSZ}\hat{C}^{-1}T^{\rm theory}_{\rm pkSZ}}\ .
\ea
We define the signal-to-noise ratio of the kSZ detection as
\ba
	\frac{\rm S}{\rm N}\equiv \frac{\bar{\tau}_e^{\rm
    bestfit}}{\sigma_{\bar\tau_e}}\ .
\ea
For the above linear fitting of a single parameter, the above definition is identical to another definition of S/N 
\ba
	\frac{\rm S}{\rm N}=\sqrt{\chi_{\rm null}^2-\chi_{\rm                    min}^2} \ .
\ea
Here $\chi^2_{\rm null}\equiv \chi^2(\bar{\tau}_e=0)$ and $\chi_{\rm min}^2\equiv \chi^2(\bar{\tau}_e^{\rm bestfit})$. 

\section{Results} \label{sec:result}
\begin{figure}
\includegraphics[width=0.5\textwidth]{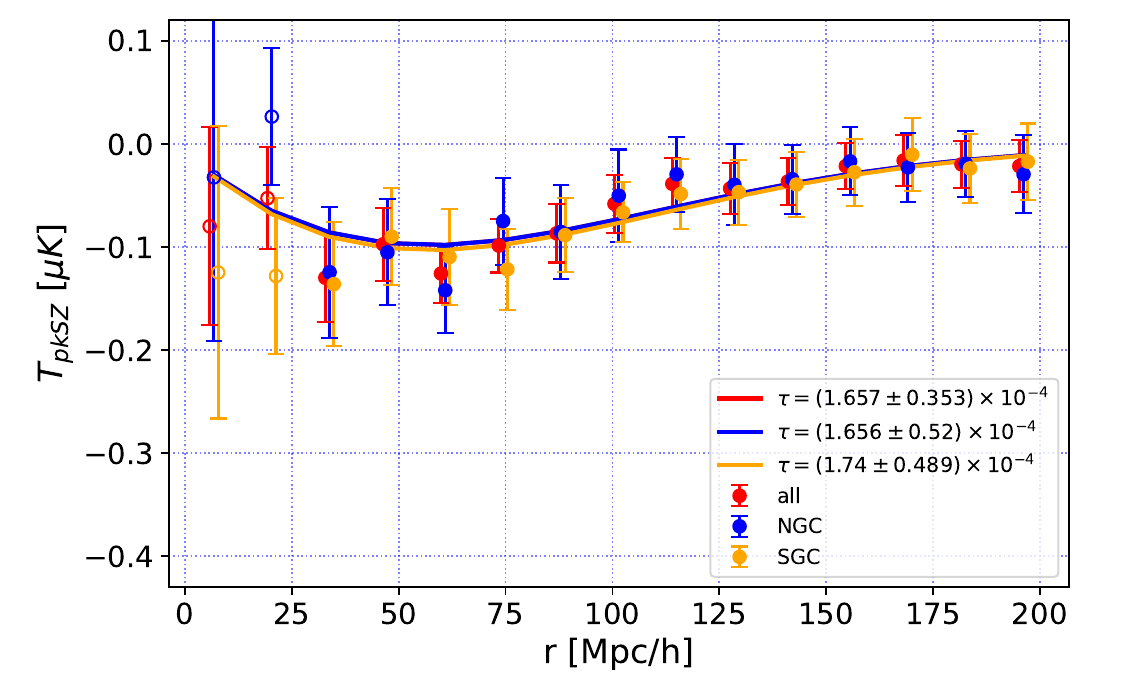}
\caption{The pairwise kSZ signal from the baseline measurement. The solid lines are the best fitting kSZ signal. The measured mean optical depth is $(1.66 \pm 0.353)\times 10^{-4}$, and S/N is 4.7. The separate measurements on the NGC and SGC cluster sub-samples are consistent with each other. \label{fig:baseline_model_pksz}}
\end{figure}

We define the baseline kSZ measurement as follows.
\bi
\item Data. The Planck 217 GHz map and the baseline cluster sample of Y21 (the richest 120,000 clusters).
\item Analysis. The kSZ theory is fixed to Eq. \ref{eq:kSZ_pairwise_2_velocity_pairwise}. For potential contaminations to be eliminated, we include both  $n_{0,1}$ (Eq. \ref{eq:model3}). The AP filter size is $3$ arcmin. The mass weight is $\propto M^{\alpha=1}$. 
\ei
We will first discuss the results of the baseline measurement. We then briefly summarize the impact of choices on Planck maps, cluster samples, noise modeling, AP filter size and mass weight. We leave the majority of the detailed analysis into the appendices.

\begin{figure}
	\centering
	\includegraphics[width=0.5\textwidth]{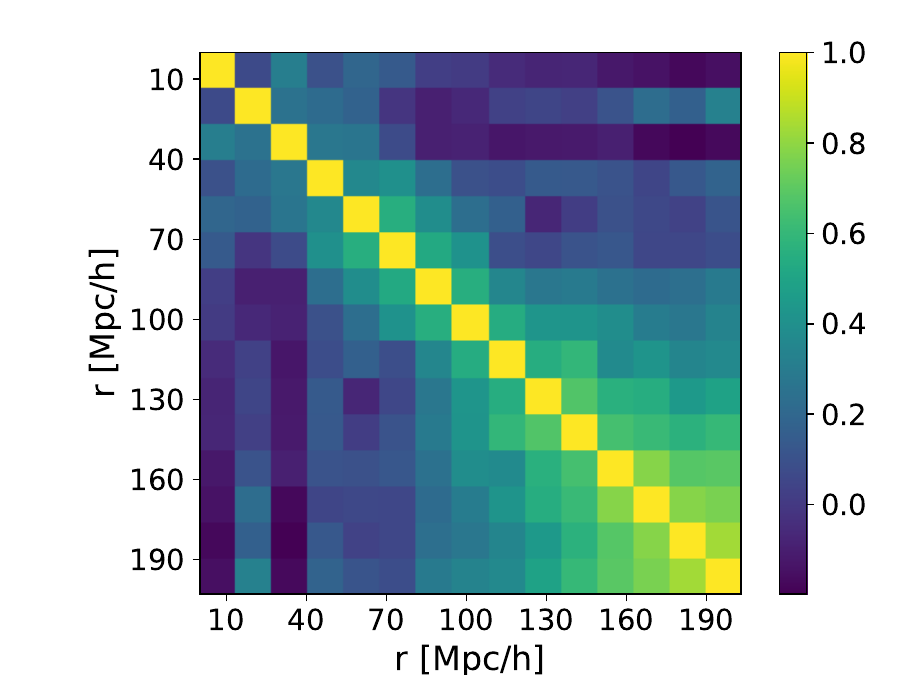}
	\caption{ The correlation matrix of the baseline measurement. The correlation between different $r$-bins become significant when $r \geq 100\ {\rm Mpc}/h. $ \label{fig:baseline_model_corva}}
\end{figure}
	
\subsection{The baseline pairwise kSZ measurement} \label{subsec:diff_model}
Fig. \ref{fig:baseline_model_pksz} shows the baseline pairwise kSZ effect measurement as a function of pair separation $r$. The pairwise kSZ signal peaks at $r \sim 50\ {\rm Mpc}/h$, with the peak amplitude $\sim -0.1\mu$K. The measurements agree well with the kSZ template from simulation (Fig. \ref{fig:baseline_model_pksz}), with S/N$=4.7$.  The mean optical depth, for the baseline cluster sample with mean mass $1.52 \times 10^{14} {\rm M_{\odot}}/h$, is
\ba
	\tau_e=(1.66 \pm 0.35) \times 10^{-4}\ .
\ea
In the fitting, we disregard the first two data points for uncertainties in the theoretical prediction and on photo-z error estimation (Appendix \ref{app:simulation}). So the degrees of freedom (d.o.f.) is $15-2-1=12$. $\chi^2_{\rm min}=4.2$, meaning a good fit. The covariance matrix used in the above fitting is given by Jackknife resampling of $100$ sub-samples in both NGC and SGC. Given the limited Jackknife samples, we correct the inverse of the covariance matrix by  a
factor $(N_{\rm JK}-N_{\rm r-bins}-2)/(N_{\rm JK}-1)$ \citep{2007A&A...464..399H}.  Here $N_{\rm r-bins}=13$ is the size of the data vector. The normalized covariance matrix ($R_{ij}\equiv C_{{\rm JK},ij}/\sqrt{C_{{\rm JK},ii}C_{{\rm JK},jj}}$) used in this analysis is shown in Fig. \ref{fig:baseline_model_corva}.  Correlation between $r$-bins increases with increasing $r$, and becomes significant at $r \ga 100\ {\rm Mpc}/h$. The reason is that pairs of larger separations have large chance of sharing common cluster members with pairs of other separations. Such large correlation is the cause of  $\chi^2_{\rm min}/{\rm d.o.f.}\sim 4/12$, instead of $\sim 1$ expected for uncorrelated data.

For comparison, we also show the separate measurements on the NGC and SGC cluster sub-samples (Fig. \ref{fig:baseline_model_pksz}  \& Table \ref{tab:results_models_SN}).  
The two results are consistent with each other. 
\begin{table}	
\centering
\begin{tabular}{cccc|cc}
\hline
	\hline
	& $S/N$ & $\chi^2_{\rm min}$ &$\bar{\tau}_e\times 10^4$ &$n_0$[$10^{-2} \mu{\rm K}$]\\
	\hline
	all & 4.69 & 3.91 & $1.66\pm 0.35$&  $1.6\pm4.5$\\
    NGC & 3.18 & 4.09  & $1.66\pm 0.52$&  $6.3\pm5.1$ \\
    SGC & 3.56 & 2.38  &  $1.74\pm 0.49$& $-4.9\pm5.1$ \\
	\hline
	\end{tabular}	
	\caption{The baseline measurements. The degrees of freedom is $12$. $n_0$ describe the major contamination. \label{tab:results_models_SN}}	
\end{table}

\begin{figure*}
	\centering
	\includegraphics[width=0.8\textwidth]{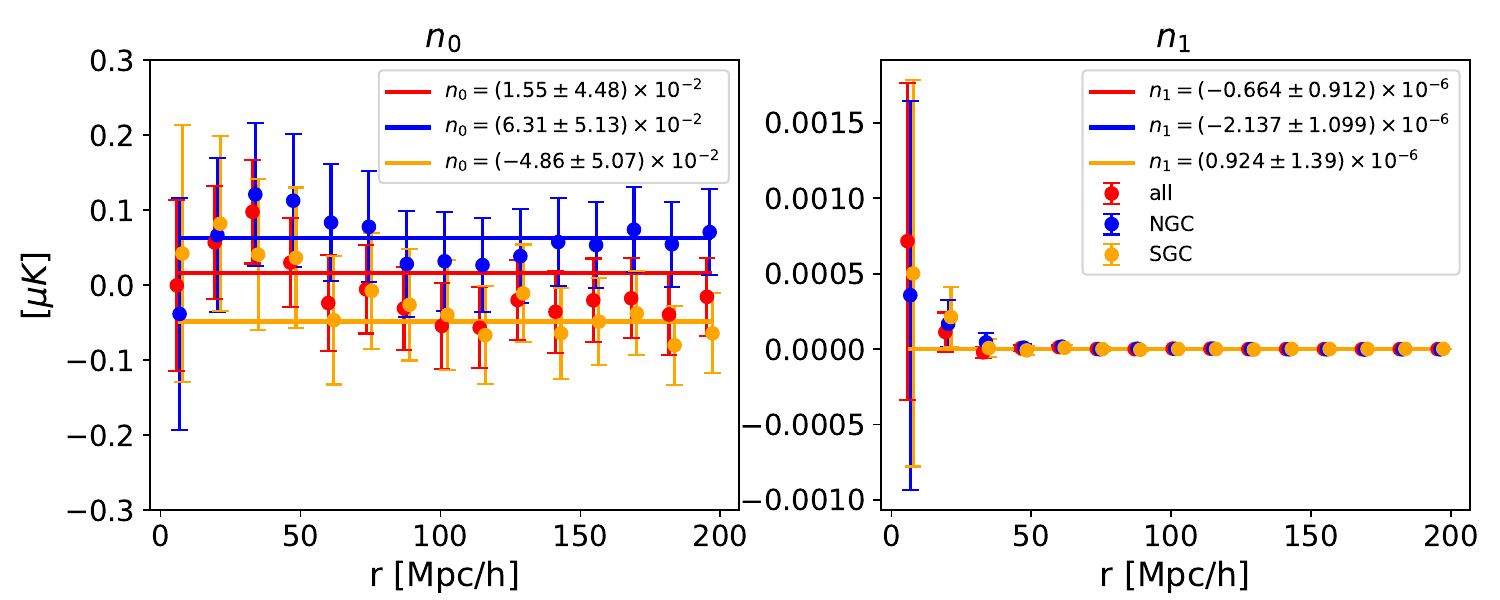}
	\caption{The potential systematic noise constraint from the baseline measurement. The points with errorbars are the measurements results, and the dashed lines are best fitting with a scale-independent model. The left panel is the redshift-independent term $n_0$. At $r \sim 30 {\rm Mpc}/h$, it is not consistent with a null signal in $1-\sigma$ region. The right panel is the redshift-dependent term $n_1$, whose amplitude is much smaller than that of $n_0$. \label{fig:baseline_model_noise}}
\end{figure*}

\subsubsection{Measures of potential systematics}
Our estimator (Eq. \ref{eq:baseline_model}) also provides estimation on potential systematic contaminations, in terms of $n_{0,1}$. Fig. \ref{fig:baseline_model_noise} and Table \ref {tab:results_models_SN} show the constraints on $n_{0,1}$. $n_0$ can be generated by various sources such as primary CMB and CIB, since we only sample their contaminations around galaxy clusters. What we find is that, $n_0$ has no significant scale dependence. Its average value is consistent with zero ($\bar{n}_0=(1.6\pm4.5)\times 10^{-2}\mu$K) for the baseline cluster sample. 

However, $n_0$ has large fluctuations across sky. Difference in $\bar{n}_0$ of the NGC and SGC sub-samples amounts to $1.6\sigma$. Also, both the $n_0$ of NGC and SGC show visible scale dependences, although it may be caused by statistical fluctuations associated with their relatively large error bars. Furthermore, $n_0$ shows strong dependence on the CMB maps used and its deviation from zero can be significant for some of the maps (appendix \ref{app:different_cmb_maps}). Together with its amplitude comparable to the kSZ signal, we should include $n_0$ in the data analysis to avoid its potentially significant contaminations. Our estimation automatically takes such kind of potential contaminations into account and safely eliminates the systematic bias induced to the kSZ measurement. 

In contrast, $n_1$ is constrained to the level of $10^{-6}\mu {\rm K}$, much weaker than the kSZ signal. So this type of contamination is completely negligible. 

Beside considering $n_0$ and $n_1$ as sources of contamination together (baseline model), we consider them separately as well (Appendix \ref{app:different_models}). Model I only considers $n_0$. The results are almost identical to the case of considering both $n_0$ and $n_1$. This is consistent with our previous finding that $n_1\lesssim 10^{-6}\mu $K, negligible comparing to the kSZ signal. Model II only considers $n_1$. The constrained $\tau$ shows significant changes comparing to both the baseline model and model I, especially for NGC and SGC.  The conclusion is that we must include $n_0$ in the analysis, but we may ignore $n_1$ in the analysis.

\subsubsection{Validation against mass weight, CMB maps, and covariance
  matrix}
To further validate the measurement, we carry out measurements with the same baseline cluster sample, but with different mass weight, different CMB maps, and covariance matrix. 

\begin{figure}
	\includegraphics[width=0.45\textwidth]{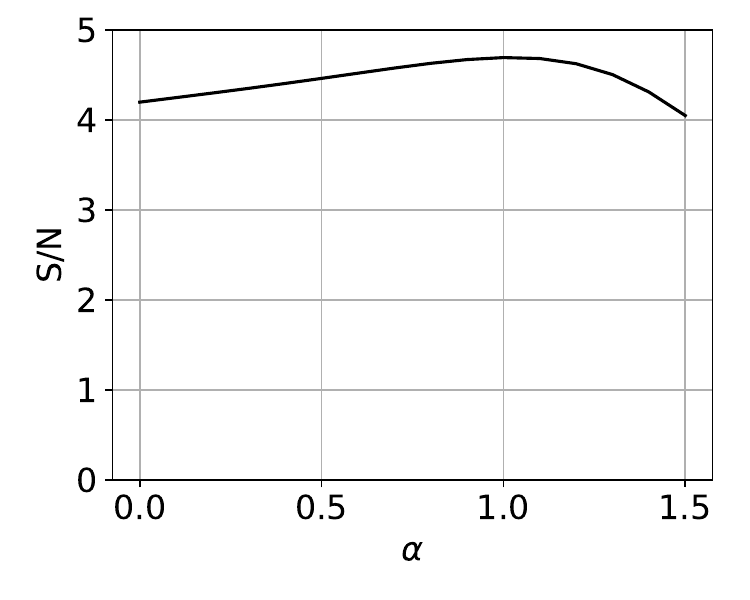}
	\caption{The dependence of S/N on the mass weight index $\alpha$ ($W\propto M^\alpha$). The S/N peaks at $\alpha\sim 1$, consistent with our expectation that the kSZ contribution is proportional to the cluster mass $M$.\label{fig:results_mass_weight_sn}}
\end{figure}

 If the measured quantity is indeed the kSZ signal, we expect that a mass weight $\propto M^{\alpha=1}$ would return the optimal measurement. Fig.  \ref{fig:results_mass_weight_sn} shows the S/N as a function of $\alpha$. It indeed peaks at $\alpha\simeq 1$, supporting the kSZ origin of the measurement. 

We also detect the kSZ effect in four foreground-cleaned CMB maps (SMICA, SEVEM, NLIC and COMMANDER, Appendix \ref{app:different_cmb_maps}). As expected, the Planck 217 GHz map  produces the highest S/N.  The  217 GHz map has the highest angular resolution, so the kSZ effect can be better separated from primary CMB by the AP filter. We also find that $n_{0}$ of the 4 foreground-cleaned  maps deviates from zero at $2$-$4\sigma$, at the first $r$ bin around $10h^{-1}$ Mpc.  This is likely caused by residual tSZ in these maps.

 We vary the number of jackknife samples to check the stability of the S/N, and find that the choice of $100$ is appropriate (Appendix \ref{app:consistency_SN}. We also perform the singular value decomposition (SVD) and confirm that the inverse of the covariance matrix is also stable. 
 
\subsubsection{Convert $\bar \tau_e$ into baryon abundance}
\begin{figure}
	\includegraphics[width=0.45\textwidth]{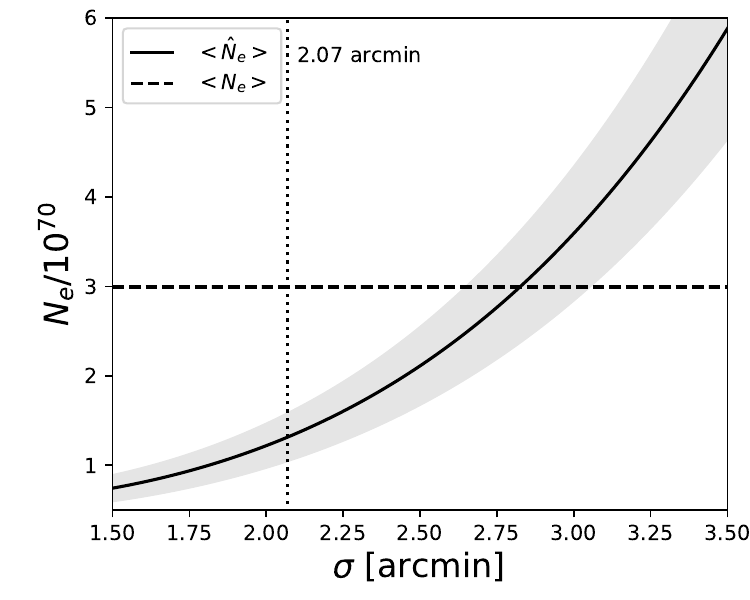}
	\caption{The measured electron number of a cluster $<\hat N_e>$ and the predicted one $< N_e>$ of the baseline sample. The shaded region is the given by the uncertainty of the optical depth and be largely underestimated. The vertical dotted line is $\sigma=\sigma_{beam}$, which represent the lower limit of the measured electron number. \label{fig:tau2electron}}
\end{figure}	

In this section, we show a rough estimate of baryon abundance by converting the measured optical depth into the mean electron number.
In the first, we need to correct the suppression caused the large beam size of Planck. 
Ideally, the size of AP filter should be as large as the virial radius. Therefore, the primary CMB contamination would be canceled out and the kSZ signal would be retained. Due to the beam effect, however, kSZ signal of clusters is smoothed. It would get to the outer region of the AP filter and then be subtracted. To characterize this effect, it can be simply assumed the profile of the smoothed optical depth is 
\ba\label{tau_profile}
	\tau (\theta)=\frac{\tau_0}{2\pi\sigma^2} e^{-\frac{\theta^2}{2\sigma^2}},
\ea
where $\sigma^2=\sigma_{beam}^2+\sigma_{eff}^2$. Then the fraction of kSZ signal in the inner region is 
\ba
	f1=\int^{\theta_{AP}}_{0}\frac{1}{2\pi\sigma^2} e^{-\frac{\theta^2}{2\sigma^2}}\times 2\pi\theta d\theta,
\ea
and the fraction in the outter region is 
\ba
	f2=\int^{\sqrt 2 \theta_{AP}}_{\theta_{AP}}\frac{1}{2\pi\sigma^2} e^{-\frac{\theta^2}{2\sigma^2}}\times 2\pi\theta d\theta.
\ea
The radio of the measured kSZ signal and the true one is 
\ba
	f_{AP}=f1-f2=1-2e^{\frac{\theta_{AP}}{2\sigma^2}}+e^{\frac{\theta_{AP}}{\sigma^2}} .
\ea
Ignoring the size of cluster ($\sigma^2=\sigma_{beam}^2$=$2.07'^2$), $f$ equals 0.37 which means we only measure 37\% kSZ signal with AP filter, $\theta_{AP}=3'$. Then we can convert $\bar \tau_e$ into electron number 
\ba
<\hat N_e>=\frac{\bar\tau_e}{f_{AP}\sigma_T}\pi \theta_{AP}^2<D_A^2>,
\ea
where $D_A$ is the angular diameter distance. The predicted mean electron number in a cluster is 
\ba
	<N_e>=\frac{<M>}{m_H}\cdot\frac{1+X_{H}}{2}\cdot\frac{\Omega_b}{\Omega_m},
\ea
where $X_H=0.76$ is the mass fraction of hydrogen. Then we compare $<\hat N_e>$ and $< N_e>$ of the baseline sample in Fig.\ref{fig:tau2electron} as a function of $\sigma$ in Eq.\ref{tau_profile}. When $\sigma=\sigma_{beam}$, the measured $<\hat N_e>$ is $\sim 50 \%$ of the predicted $<N_e>$. Including the size of cluster, $\sigma$ should be larger and correspending to a larger $<\hat N_e>$. Roughly speaking, we have measured all baryons in clusters of the baseline sample.

\subsection{Other cluster samples}
\begin{figure}
 	\centering
   	\includegraphics[width=0.5\textwidth]{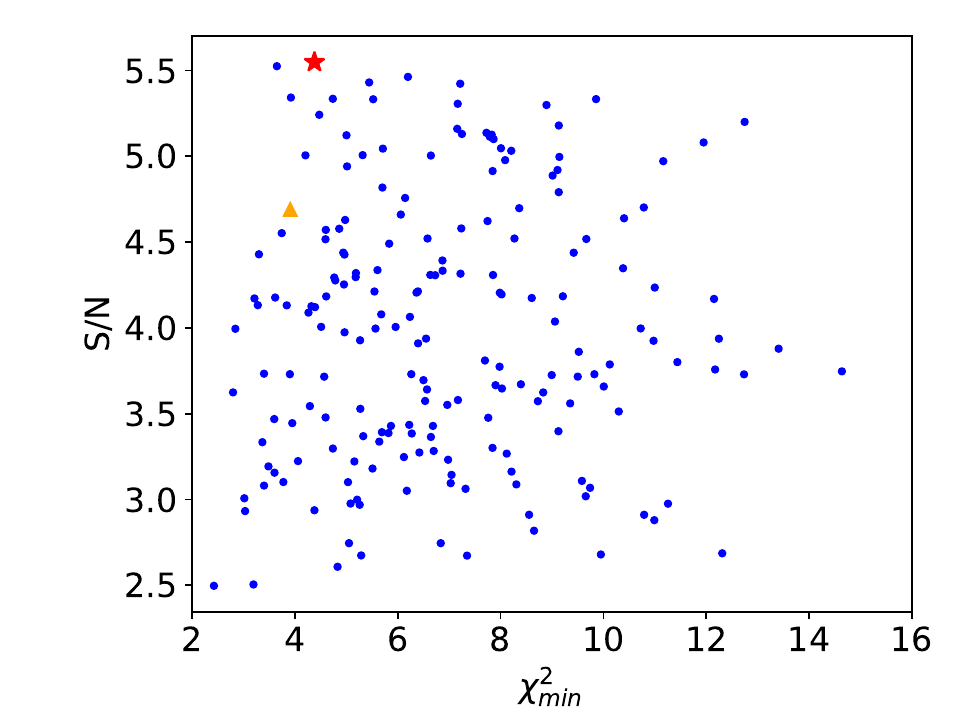}
   	\caption{The S/N and $\chi^2_{\rm min}$ of cluster samples with various mass and angular radius threshold. The red star point is the highest S/N among tested cluster samples, whose S/N=5.55 and $\chi^2_{\rm min}=4.38$. The orange triangle point represents the baseline sample. The measurements of these samples are consistent of each other, and some of them have higher S/N than that of baseline measurement.\label{fig:results_chi2_sn}}
\end{figure}
The S/N can be further improved by better defined cluster samples, and by varying AP filter sizes. This is explored in the Appendix \ref{app:choices_of_galaxy_sample}. The resulting S/N is shown in Fig.\ref{fig:APP_sample_sn}. The kSZ effect is consistently detected in most samples, with acceptable $\chi^2_{\rm min}/$d.o.f.  Fig. \ref{fig:results_chi2_sn} summarizes the S/N and $\chi^2_{\rm min}$ of these cluster samples.  Some samples show improvements in S/N over the baseline measurement.  The optimal $\theta_{\rm AP}$ varies with mass and angular size cut. But $\theta_{\rm AP}\sim 3.5^{'}$ is a good choice for all cluster samples investigated. 
\subsubsection{The cluster sample of highest S/N}
\begin{figure}
   	\centering
	\includegraphics[width=0.45\textwidth]{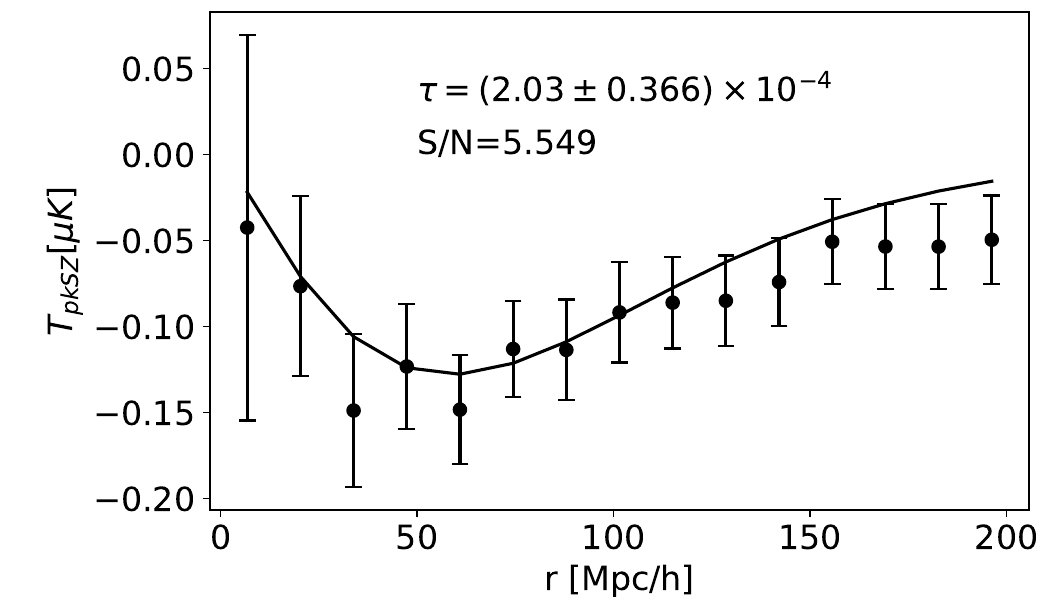}
   	\caption{The highest S/N kSZ pairwise measurement ($5.5\sigma$) is achieved with clusters of $M>10^{13.8}\ M_{\odot}$ and angular radius larger than $3.0^{'}$. $\bar{\tau}_e=(2.03 \pm 0.37) \times 10^{-4}$.\label{fig:pksz_highest_sn}}
\end{figure}

The highest S/N=$5.55$ is achieved for the cluster sample shown as the star point shown in Fig. \ref{fig:pksz_highest_sn}. This sample has a mass cut $M>10^{13.8} M_{\sun}$ and angular virial radius cut $3.0^{'}$. The first criterion selects clusters with intrinsically larger kSZ, while the second criterion selects clusters less contaminated by primary CMB. The optical depth is constrained to $(2.03 \pm 0.37) \times 10^{-4}$. 

\begin{figure}
   	\centering
   	\includegraphics[width=0.48\textwidth]{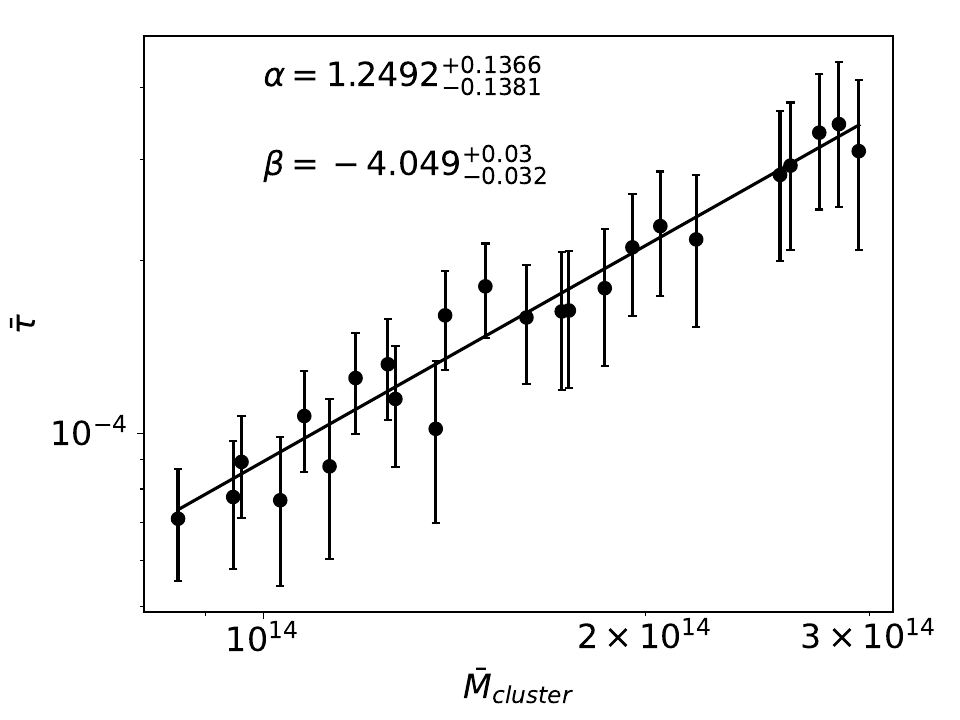}
   	\caption{The relationship between optical depth and cluster mass. The points are the measured mean optical depth of cluster sample in Fig. \ref{fig:APP_sample_sn} with $\theta_{\rm AP}$ fixing to 3.4'. The solid line is the fitting result of Eq. \ref{t_M}. The scaling is consistent with $\tau \propto M$.\label{fig:results_tau_M}}
\end{figure}
\subsubsection{The mass-$\tau$ relation}\label{subsec:mass_tau}
One further support that we consistently detect the kSZ signal over the cluster samples is Fig. \ref{fig:results_tau_M}. This figure plots the mean optical depth $\bar{\tau}_e$ versus the mean cluster mass $\bar{M}$, of various cluster samples. The AP filter size is fixed to $\theta_{\rm AP}=3.4^{'}$ so that contribution to $\bar{\tau}_e$ of different cluster samples roughly arises from the same regions around clusters. Fig. \ref{fig:results_tau_M} shows a clear trend of increasing $\bar{\tau}_e$ with $\bar{M}$. It can be well fitted with a power-law,
\ba
	\label{t_M}
    \lg \bar{\tau}_e = \alpha(\lg \bar{M} -14) + \beta\ .
\ea
We find
\ba
\alpha=1.25 \pm 0.14\ .
\ea
Therefore the $M$-$\tau$ scaling is consistent with $\tau\propto M$, as expected for massive clusters. Nevertheless, we caution that the verification of $\tau\propto M$ is very rough, for two reasons. Firstly, these cluster samples are not independent of each other, so the uncertainty in $\alpha$ is underestimated. Secondly, their redshift distributions are not identical to each other.  Due to the limited S/N, we have to combine all clusters over a wide range of  redshift. Therefore mass or angular size cut both alter the redshift distribution. The situation will be improved with future high resolution and high sensitivity CMB data, with which we can choose sufficiently narrow redshift bins and split clusters into separate mass bins. 

\subsubsection{The cluster sample with spectroscopic redshift} \label{app:spec}
For pairwise kSZ measurement, it is important to know which member  of a cluster pair is closer to the observer. The large uncertainty of photometric redshift may lead to wrong distance order of cluster pairs. This can reduce the kSZ signal by a factor of $\sim 3$ (Fig. \ref{fig:app_simu_v12} \& \ref{fig:app_simu_v12_zph}). Therefore if we have spectroscopic redshift measurements of all clusters, we expect a total S/N $\sim 15$. The ongoing DESI experiment will be able to measure spectroscopic redshifts for at least a fraction of these clusters, and will then improve the kSZ measurement. 
	\begin{figure}
	\centering
	\includegraphics[width=0.5\textwidth]{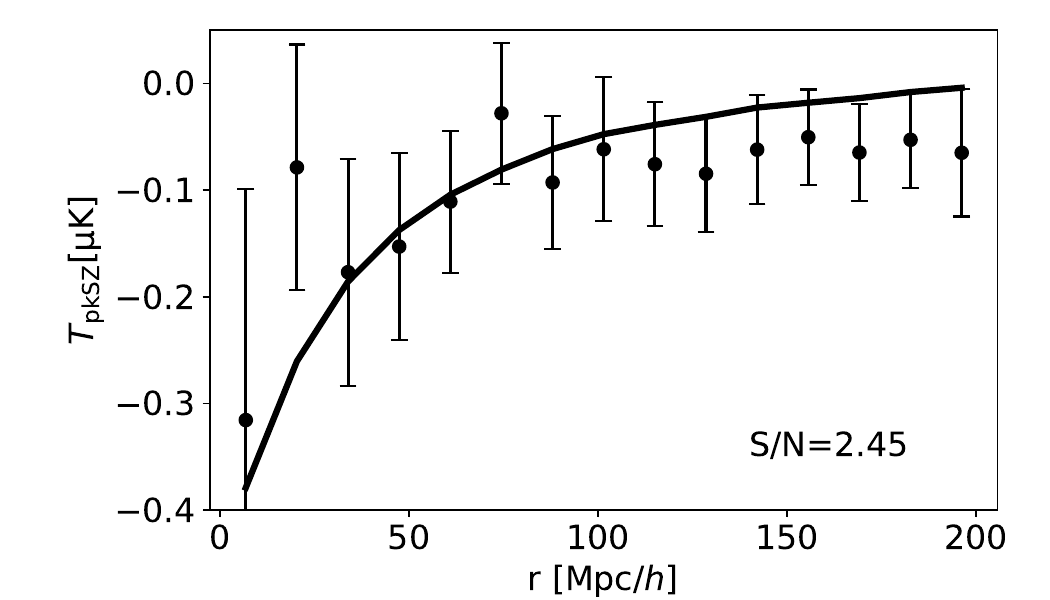}
 	\caption{The pairwise kSZ measured by a spectroscopic redshift sample of $33,000$ clusters. The optimal fitting of optical depth is $(0.73\pm0.3)\times 10^{-4}$. This value is lower than the result of the baseline model. Because these two samples are not totally same. In addition, the mis-position of clusters and their BCG and the redshift uncertainty caused by peculiar velocity may result in the same effect as photo-z but they are not included in the simulated template. And this may cause the template higher than it should be and lower the fitting optical depth.\label{fig:app_specz}}
	\end{figure}

Fortunately, some clusters in the Y12 catalog have spectroscopic redshift information from pre-existing surveys. Most of them are in NGC. Among 60,000 NGC clusters in our baseline sample, $\sim 33,000$ clusters have a brightest central galaxy (BCG) member with spectroscopic redshift. The kSZ measurement is shown in Fig. \ref{fig:app_specz}.  Since the S/N scales as $N^{1/2}_{\rm pair}\propto N_{\rm cluster}$, we would expect S/N$\sim 1.7$. But the actual S/N is 2.45 and the improvement shows the contribution from spectroscopic redshift. With future DESI redshifts, we will further explore this issue. 

\section{Conclusion} \label{sec:conclu}
In this work, we measure the pairwise kSZ combining the Y12 group catalog and Planck maps, with a new kSZ estimator. We find that the Planck 217 GHz intensity map is most suitable for this purpose. The Y12 catalog provides a variety of cluster samples, and enables us to detect the kSZ effect with S/N$\simeq 5$.  For the baseline cluster sample of mean mass $1.5\times 10^{14} M_\odot/h$, we find the mean optical depth $\bar{\tau}_e=(1.66\pm 0.35)\times 10^{-4}$. We also perform a series of tests to verify the kSZ measurement. We confirm that the measured signal has the expected mass dependence and scale $r$ dependence. Our pairwise kSZ estimator also diagnoses and eliminates potential contaminations. We find that the major contamination behaves like a constant term $n_0$. Its amplitude and detection significance vary with cluster samples, pair separation and CMB maps, but the typical amplitude is $\mathcal{O}(10^{-2})\mu$K. Comparing to the peak kSZ signal of $\sim 0.1\mu$K, this $n_0$ type of contamination should be taken care of in the data analysis. 

The above kSZ measurement can be significantly improved by higher angular resolution CMB experiments such as CMB-S4, and/or  spectroscopic redshift measurement of clusters with DESI or other galaxy surveys. Then the kSZ effect can be used to constrain the baryon content in clusters and its evolution with redshift. We caution that the current template adopted in our analysis and in many previous works is only sufficient for robust evaluation of the detection S/N. It is not sufficient for accurate interpretation of the signal, in particular for comparison between clusters of different mass, redshift, AP filter size and berween different CMB maps of  different angular resolution and noise.   With significant improvement of the kSZ measurement in the near future, we shall also improve the theoretical template to interpret the measurement accurately. 

\section*{acknowledgments}
We thank Yipeng Jing for useful suggestions to analyze the cluster sample with spectroscopic redshifts. This work made use of the Gravity Supercomputer at the Department of Astronomy, Shanghai Jiao Tong University. This work was supported by the National Science Foundation of China 11621303 \& 11653003, National key R\&D Program of China (Grant No.2020 YFC2201602) and CSST CMS-CSST-2021-A02. 

\section*{data availability}
The data underlying this article will be shared on reasonable request to the corresponding author.

\bibliography{pairwise_ksz}{}

\begin{thebibliography}{}
\expandafter\ifx\csname natexlab\endcsname\relax\def\natexlab#1{#1}\fi
\providecommand{\url}[1]{\href{#1}{#1}}
\providecommand{\dodoi}[1]{doi:~\href{http://doi.org/#1}{\nolinkurl{#1}}}
\providecommand{\doeprint}[1]{\href{http://ascl.net/#1}{\nolinkurl{http://ascl.net/#1}}}
\providecommand{\doarXiv}[1]{\href{https://arxiv.org/abs/#1}{\nolinkurl{https://arxiv.org/abs/#1}}}

\bibitem[{{Adam} {et~al.}(2017){Adam}, {Bartalucci}, {Pratt}, {Ade},
  {Andr{\'e}}, {Arnaud}, {Beelen}, {Beno{\^\i}t}, {Bideaud}, {Billot},
  {Bourdin}, {Bourrion}, {Calvo}, {Catalano}, {Coiffard}, {Comis}, {D'Addabbo},
  {De Petris}, {D{\'e}mocl{\`e}s}, {D{\'e}sert}, {Doyle}, {Egami}, {Ferrari},
  {Goupy}, {Kramer}, {Lagache}, {Leclercq}, {Mac{\'\i}as-P{\'e}rez},
  {Maurogordato}, {Mauskopf}, {Mayet}, {Monfardini}, {Mroczkowski}, {Pajot},
  {Pascale}, {Perotto}, {Pisano}, {Pointecouteau}, {Ponthieu}, {Rev{\'e}ret},
  {Ritacco}, {Rodriguez}, {Romero}, {Ruppin}, {Schuster}, {Sievers},
  {Triqueneaux}, {Tucker}, {Zemcov}, \& {Zylka}}]{2017A&A...598A.115A}
{Adam}, R., {Bartalucci}, I., {Pratt}, G.~W., {et~al.} 2017, \aap, 598, A115,
  \dodoi{10.1051/0004-6361/201629182}

\bibitem[{{Aghanim} {et~al.}(2008){Aghanim}, {Majumdar}, \&
  {Silk}}]{2008RPPh...71f6902A}
{Aghanim}, N., {Majumdar}, S., \& {Silk}, J. 2008, Reports on Progress in
  Physics, 71, 066902, \dodoi{10.1088/0034-4885/71/6/066902}

\bibitem[{{Alvarez}(2016)}]{2016ApJ...824..118A}
{Alvarez}, M.~A. 2016, \apj, 824, 118, \dodoi{10.3847/0004-637X/824/2/118}

\bibitem[{{Amodeo} {et~al.}(2021){Amodeo}, {Battaglia}, {Schaan}, {Ferraro},
  {Moser}, {Aiola}, {Austermann}, {Beall}, {Bean}, {Becker}, {Bond},
  {Calabrese}, {Calafut}, {Choi}, {Denison}, {Devlin}, {Duff}, {Duivenvoorden},
  {Dunkley}, {D{\"u}nner}, {Gallardo}, {Hall}, {Han}, {Hill}, {Hilton},
  {Hilton}, {Hlo{\v{z}}ek}, {Hubmayr}, {Huffenberger}, {Hughes}, {Koopman},
  {MacInnis}, {McMahon}, {Madhavacheril}, {Moodley}, {Mroczkowski}, {Naess},
  {Nati}, {Newburgh}, {Niemack}, {Page}, {Partridge}, {Schillaci}, {Sehgal},
  {Sif{\'o}n}, {Spergel}, {Staggs}, {Storer}, {Ullom}, {Vale}, {van Engelen},
  {Van Lanen}, {Vavagiakis}, {Wollack}, \& {Xu}}]{2021PhRvD.103f3514A}
{Amodeo}, S., {Battaglia}, N., {Schaan}, E., {et~al.} 2021, \prd, 103, 063514,
  \dodoi{10.1103/PhysRevD.103.063514}

\bibitem[{{Battaglia} {et~al.}(2017){Battaglia}, {Ferraro}, {Schaan}, \&
  {Spergel}}]{2017JCAP...11..040B}
{Battaglia}, N., {Ferraro}, S., {Schaan}, E., \& {Spergel}, D.~N. 2017, \jcap,
  2017, 040, \dodoi{10.1088/1475-7516/2017/11/040}

\bibitem[{{Battaglia} {et~al.}(2013){Battaglia}, {Natarajan}, {Trac}, {Cen}, \&
  {Loeb}}]{2013ApJ...776...83B}
{Battaglia}, N., {Natarajan}, A., {Trac}, H., {Cen}, R., \& {Loeb}, A. 2013,
  \apj, 776, 83, \dodoi{10.1088/0004-637X/776/2/83}

\bibitem[{{Bhattacharya} \& {Kosowsky}(2008)}]{2008PhRvD..77h3004B}
{Bhattacharya}, S., \& {Kosowsky}, A. 2008, \prd, 77, 083004,
  \dodoi{10.1103/PhysRevD.77.083004}

\bibitem[{{Birkinshaw}(1999)}]{1999PhR...310...97B}
{Birkinshaw}, M. 1999, \physrep, 310, 97, \dodoi{10.1016/S0370-1573(98)00080-5}

\bibitem[{{Calafut} {et~al.}(2021){Calafut}, {Gallardo}, {Vavagiakis},
  {Amodeo}, {Aiola}, {Austermann}, {Battaglia}, {Battistelli}, {Beall}, {Bean},
  {Bond}, {Calabrese}, {Choi}, {Cothard}, {Devlin}, {Duell}, {Duivenvoorden},
  {Dunkley}, {Dunner}, {Ferraro}, {Guan}, {Hill}, {Hilton}, {Hlozek}, {Huber},
  {Hubmayr}, {Huffenberger}, {Hughes}, {Koopman}, {Kosowsky}, {Li}, {Lokken},
  {Madhavacheril}, {McMahon}, {Moodley}, {Naess}, {Nati}, {Newburgh},
  {Niemack}, {Partridge}, {Schaan}, {Schillaci}, {Sifon}, {Spergel}, {Staggs},
  {Ullom}, {Vale}, {Van Engelen}, {Wollack}, \& {Xu}}]{2021arXiv210108374C}
{Calafut}, V., {Gallardo}, P.~A., {Vavagiakis}, E.~M., {et~al.} 2021, arXiv
  e-prints, arXiv:2101.08374.
\newblock \doarXiv{2101.08374}

\bibitem[{{Carlstrom} {et~al.}(2002){Carlstrom}, {Holder}, \&
  {Reese}}]{2002ARA&A..40..643C}
{Carlstrom}, J.~E., {Holder}, G.~P., \& {Reese}, E.~D. 2002, \araa, 40, 643,
  \dodoi{10.1146/annurev.astro.40.060401.093803}

\bibitem[{{Cayuso} \& {Johnson}(2020)}]{2020PhRvD.101l3508C}
{Cayuso}, J.~I., \& {Johnson}, M.~C. 2020, \prd, 101, 123508,
  \dodoi{10.1103/PhysRevD.101.123508}

\bibitem[{{Chaves-Montero} {et~al.}(2021){Chaves-Montero},
  {Hern{\'a}ndez-Monteagudo}, {Angulo}, \& {Emberson}}]{2021MNRAS.503.1798C}
{Chaves-Montero}, J., {Hern{\'a}ndez-Monteagudo}, C., {Angulo}, R.~E., \&
  {Emberson}, J.~D. 2021, \mnras, 503, 1798, \dodoi{10.1093/mnras/staa3782}

\bibitem[{{De Bernardis} {et~al.}(2017){De Bernardis}, {Aiola}, {Vavagiakis},
  {Battaglia}, {Niemack}, {Beall}, {Becker}, {Bond}, {Calabrese}, {Cho},
  {Coughlin}, {Datta}, {Devlin}, {Dunkley}, {Dunner}, {Ferraro}, {Fox},
  {Gallardo}, {Halpern}, {Hand}, {Hasselfield}, {Henderson}, {Hill}, {Hilton},
  {Hilton}, {Hincks}, {Hlozek}, {Hubmayr}, {Huffenberger}, {Hughes}, {Irwin},
  {Koopman}, {Kosowsky}, {Li}, {Louis}, {Lungu}, {Madhavacheril}, {Maurin},
  {McMahon}, {Moodley}, {Naess}, {Nati}, {Newburgh}, {Nibarger}, {Page},
  {Partridge}, {Schaan}, {Schmitt}, {Sehgal}, {Sievers}, {Simon}, {Spergel},
  {Staggs}, {Stevens}, {Thornton}, {van Engelen}, {Van Lanen}, \&
  {Wollack}}]{2017JCAP...03..008D}
{De Bernardis}, F., {Aiola}, S., {Vavagiakis}, E.~M., {et~al.} 2017, \jcap,
  2017, 008, \dodoi{10.1088/1475-7516/2017/03/008}

\bibitem[{{Deutsch} {et~al.}(2018){Deutsch}, {Dimastrogiovanni}, {Johnson},
  {M{\"u}nchmeyer}, \& {Terrana}}]{2018PhRvD..98l3501D}
{Deutsch}, A.-S., {Dimastrogiovanni}, E., {Johnson}, M.~C., {M{\"u}nchmeyer},
  M., \& {Terrana}, A. 2018, \prd, 98, 123501,
  \dodoi{10.1103/PhysRevD.98.123501}

\bibitem[{{Dor{\'e}} {et~al.}(2003){Dor{\'e}}, {Knox}, \&
  {Peel}}]{2003ApJ...585L..81D}
{Dor{\'e}}, O., {Knox}, L., \& {Peel}, A. 2003, \apjl, 585, L81,
  \dodoi{10.1086/374542}

\bibitem[{{Dunkley} {et~al.}(2013){Dunkley}, {Calabrese}, {Sievers}, {Addison},
  {Battaglia}, {Battistelli}, {Bond}, {Das}, {Devlin}, {D{\"u}nner}, {Fowler},
  {Gralla}, {Hajian}, {Halpern}, {Hasselfield}, {Hincks}, {Hlozek}, {Hughes},
  {Irwin}, {Kosowsky}, {Louis}, {Marriage}, {Marsden}, {Menanteau}, {Moodley},
  {Niemack}, {Nolta}, {Page}, {Partridge}, {Sehgal}, {Spergel}, {Staggs},
  {Switzer}, {Trac}, \& {Wollack}}]{2013JCAP...07..025D}
{Dunkley}, J., {Calabrese}, E., {Sievers}, J., {et~al.} 2013, \jcap, 2013, 025,
  \dodoi{10.1088/1475-7516/2013/07/025}

\bibitem[{{Fan} {et~al.}(2006){Fan}, {Carilli}, \&
  {Keating}}]{2006ARA&A..44..415F}
{Fan}, X., {Carilli}, C.~L., \& {Keating}, B. 2006, \araa, 44, 415,
  \dodoi{10.1146/annurev.astro.44.051905.092514}

\bibitem[{{George} {et~al.}(2015){George}, {Reichardt}, {Aird}, {Benson},
  {Bleem}, {Carlstrom}, {Chang}, {Cho}, {Crawford}, {Crites}, {de Haan},
  {Dobbs}, {Dudley}, {Halverson}, {Harrington}, {Holder}, {Holzapfel}, {Hou},
  {Hrubes}, {Keisler}, {Knox}, {Lee}, {Leitch}, {Lueker}, {Luong-Van},
  {McMahon}, {Mehl}, {Meyer}, {Millea}, {Mocanu}, {Mohr}, {Montroy}, {Padin},
  {Plagge}, {Pryke}, {Ruhl}, {Schaffer}, {Shaw}, {Shirokoff}, {Spieler},
  {Staniszewski}, {Stark}, {Story}, {van Engelen}, {Vanderlinde}, {Vieira},
  {Williamson}, \& {Zahn}}]{2015ApJ...799..177G}
{George}, E.~M., {Reichardt}, C.~L., {Aird}, K.~A., {et~al.} 2015, \apj, 799,
  177, \dodoi{10.1088/0004-637X/799/2/177}

\bibitem[{{Goodman}(1995)}]{1995PhRvD..52.1821G}
{Goodman}, J. 1995, \prd, 52, 1821, \dodoi{10.1103/PhysRevD.52.1821}

\bibitem[{{G{\'o}rski} {et~al.}(2005){G{\'o}rski}, {Hivon}, {Banday},
  {Wandelt}, {Hansen}, {Reinecke}, \& {Bartelmann}}]{2005ApJ...622..759G}
{G{\'o}rski}, K.~M., {Hivon}, E., {Banday}, A.~J., {et~al.} 2005, \apj, 622,
  759, \dodoi{10.1086/427976}

\bibitem[{{Hand} {et~al.}(2012){Hand}, {Addison}, {Aubourg}, {Battaglia},
  {Battistelli}, {Bizyaev}, {Bond}, {Brewington}, {Brinkmann}, {Brown}, {Das},
  {Dawson}, {Devlin}, {Dunkley}, {Dunner}, {Eisenstein}, {Fowler}, {Gralla},
  {Hajian}, {Halpern}, {Hilton}, {Hincks}, {Hlozek}, {Hughes}, {Infante},
  {Irwin}, {Kosowsky}, {Lin}, {Malanushenko}, {Malanushenko}, {Marriage},
  {Marsden}, {Menanteau}, {Moodley}, {Niemack}, {Nolta}, {Oravetz}, {Page},
  {Palanque-Delabrouille}, {Pan}, {Reese}, {Schlegel}, {Schneider}, {Sehgal},
  {Shelden}, {Sievers}, {Sif{\'o}n}, {Simmons}, {Snedden}, {Spergel}, {Staggs},
  {Swetz}, {Switzer}, {Trac}, {Weaver}, {Wollack}, {Yeche}, \&
  {Zunckel}}]{2012PhRvL.109d1101H}
{Hand}, N., {Addison}, G.~E., {Aubourg}, E., {et~al.} 2012, \prl, 109, 041101,
  \dodoi{10.1103/PhysRevLett.109.041101}

\bibitem[{Hand {et~al.}(2012)Hand, Addison, Aubourg, Battaglia, Battistelli,
  Bizyaev, Bond, Brewington, Brinkmann, Brown, \& et~al.}]{Hand_2012}
Hand, N., Addison, G.~E., Aubourg, E., {et~al.} 2012, Physical Review Letters,
  109, \dodoi{10.1103/physrevlett.109.041101}

\bibitem[{{Hartlap} {et~al.}(2007){Hartlap}, {Simon}, \&
  {Schneider}}]{2007A&A...464..399H}
{Hartlap}, J., {Simon}, P., \& {Schneider}, P. 2007, \aap, 464, 399,
  \dodoi{10.1051/0004-6361:20066170}

\bibitem[{{Hern{\'a}ndez-Monteagudo} {et~al.}(2015){Hern{\'a}ndez-Monteagudo},
  {Ma}, {Kitaura}, {Wang}, {G{\'e}nova-Santos}, {Mac{\'\i}as-P{\'e}rez}, \&
  {Herranz}}]{2015PhRvL.115s1301H}
{Hern{\'a}ndez-Monteagudo}, C., {Ma}, Y.-Z., {Kitaura}, F.~S., {et~al.} 2015,
  \prl, 115, 191301, \dodoi{10.1103/PhysRevLett.115.191301}

\bibitem[{{Hill} {et~al.}(2016){Hill}, {Ferraro}, {Battaglia}, {Liu}, \&
  {Spergel}}]{2016PhRvL.117e1301H}
{Hill}, J.~C., {Ferraro}, S., {Battaglia}, N., {Liu}, J., \& {Spergel}, D.~N.
  2016, \prl, 117, 051301, \dodoi{10.1103/PhysRevLett.117.051301}

\bibitem[{{Hoscheit} \& {Barger}(2018)}]{2018ApJ...854...46H}
{Hoscheit}, B.~L., \& {Barger}, A.~J. 2018, \apj, 854, 46,
  \dodoi{10.3847/1538-4357/aaa59b}

\bibitem[{{Hotinli} \& {Johnson}(2020)}]{2020arXiv201209851H}
{Hotinli}, S.~C., \& {Johnson}, M.~C. 2020, arXiv e-prints, arXiv:2012.09851.
\newblock \doarXiv{2012.09851}

\bibitem[{{Hu}(2000)}]{2000ApJ...529...12H}
{Hu}, W. 2000, \apj, 529, 12, \dodoi{10.1086/308279}

\bibitem[{{Iliev} {et~al.}(2007){Iliev}, {Pen}, {Bond}, {Mellema}, \&
  {Shapiro}}]{2007ApJ...660..933I}
{Iliev}, I.~T., {Pen}, U.-L., {Bond}, J.~R., {Mellema}, G., \& {Shapiro}, P.~R.
  2007, \apj, 660, 933, \dodoi{10.1086/513687}

\bibitem[{{Iliev} {et~al.}(2006){Iliev}, {Pen}, {Richard Bond}, {Mellema}, \&
  {Shapiro}}]{2006NewAR..50..909I}
{Iliev}, I.~T., {Pen}, U.-L., {Richard Bond}, J., {Mellema}, G., \& {Shapiro},
  P.~R. 2006, \nar, 50, 909, \dodoi{10.1016/j.newar.2006.09.012}

\bibitem[{{Jimenez} {et~al.}(2019){Jimenez}, {Maartens}, {Rida Khalifeh},
  {Caldwell}, {Heavens}, \& {Verde}}]{2019JCAP...05..048J}
{Jimenez}, R., {Maartens}, R., {Rida Khalifeh}, A., {et~al.} 2019, \jcap, 2019,
  048, \dodoi{10.1088/1475-7516/2019/05/048}

\bibitem[{Jing(2018)}]{Jing_2018}
Jing, Y. 2018, Science China Physics, Mechanics \& Astronomy, 62,
  \dodoi{10.1007/s11433-018-9286-x}

\bibitem[{{Li} {et~al.}(2018){Li}, {Ma}, {Remazeilles}, \&
  {Moodley}}]{2018PhRvD..97b3514L}
{Li}, Y.-C., {Ma}, Y.-Z., {Remazeilles}, M., \& {Moodley}, K. 2018, \prd, 97,
  023514, \dodoi{10.1103/PhysRevD.97.023514}

\bibitem[{{Lim} {et~al.}(2020){Lim}, {Mo}, {Wang}, \&
  {Yang}}]{2020ApJ...889...48L}
{Lim}, S.~H., {Mo}, H.~J., {Wang}, H., \& {Yang}, X. 2020, \apj, 889, 48,
  \dodoi{10.3847/1538-4357/ab63df}

\bibitem[{{Ma} \& {Zhao}(2014)}]{2014PhLB..735..402M}
{Ma}, Y.-Z., \& {Zhao}, G.-B. 2014, Physics Letters B, 735, 402,
  \dodoi{10.1016/j.physletb.2014.06.066}

\bibitem[{{McQuinn} {et~al.}(2005){McQuinn}, {Furlanetto}, {Hernquist}, {Zahn},
  \& {Zaldarriaga}}]{2005ApJ...630..643M}
{McQuinn}, M., {Furlanetto}, S.~R., {Hernquist}, L., {Zahn}, O., \&
  {Zaldarriaga}, M. 2005, \apj, 630, 643, \dodoi{10.1086/432049}

\bibitem[{{Mesinger} {et~al.}(2012){Mesinger}, {McQuinn}, \&
  {Spergel}}]{2012MNRAS.422.1403M}
{Mesinger}, A., {McQuinn}, M., \& {Spergel}, D.~N. 2012, \mnras, 422, 1403,
  \dodoi{10.1111/j.1365-2966.2012.20713.x}

\bibitem[{{Mitchell} {et~al.}(2021){Mitchell}, {Arnold},
  {Hern{\'a}ndez-Aguayo}, \& {Li}}]{2021MNRAS.501.4565M}
{Mitchell}, M.~A., {Arnold}, C., {Hern{\'a}ndez-Aguayo}, C., \& {Li}, B. 2021,
  \mnras, 501, 4565, \dodoi{10.1093/mnras/staa3941}

\bibitem[{{Mo} {et~al.}(2010){Mo}, {van den Bosch}, \&
  {White}}]{2010gfe..book.....M}
{Mo}, H., {van den Bosch}, F.~C., \& {White}, S. 2010, {Galaxy Formation and
  Evolution}

\bibitem[{Mueller {et~al.}(2015)Mueller, Bernardis, Bean, \&
  Niemack}]{Mueller_2015}
Mueller, E.-M., Bernardis, F.~d., Bean, R., \& Niemack, M.~D. 2015, The
  Astrophysical Journal, 808, 47, \dodoi{10.1088/0004-637x/808/1/47}

\bibitem[{{Mueller} {et~al.}(2015){Mueller}, {de Bernardis}, {Bean}, \&
  {Niemack}}]{2015ApJ...808...47M}
{Mueller}, E.-M., {de Bernardis}, F., {Bean}, R., \& {Niemack}, M.~D. 2015,
  \apj, 808, 47, \dodoi{10.1088/0004-637X/808/1/47}

\bibitem[{{Munshi} {et~al.}(2016){Munshi}, {Iliev}, {Dixon}, \&
  {Coles}}]{2016MNRAS.463.2425M}
{Munshi}, D., {Iliev}, I.~T., {Dixon}, K.~L., \& {Coles}, P. 2016, \mnras, 463,
  2425, \dodoi{10.1093/mnras/stw2067}

\bibitem[{{Pen} \& {Zhang}(2014)}]{2014PhRvD..89f3009P}
{Pen}, U.-L., \& {Zhang}, P. 2014, \prd, 89, 063009,
  \dodoi{10.1103/PhysRevD.89.063009}

\bibitem[{{Planck Collaboration} {et~al.}(2014){Planck Collaboration}, {Ade},
  {Aghanim}, {Arnaud}, {Ashdown}, {Aumont}, {Baccigalupi}, {Balbi}, {Banday},
  {Barreiro}, {Battaner}, {Benabed}, {Benoit-L{\'e}vy}, {Bernard},
  {Bersanelli}, {Bielewicz}, {Bikmaev}, {Bobin}, {Bock}, {Bonaldi}, {Bond},
  {Borrill}, {Bouchet}, {Burigana}, {Butler}, {Cabella}, {Cardoso}, {Catalano},
  {Chamballu}, {Chiang}, {Chon}, {Christensen}, {Clements}, {Colombi},
  {Colombo}, {Crill}, {Cuttaia}, {Da Silva}, {Dahle}, {Davies}, {Davis}, {de
  Bernardis}, {de Gasperis}, {de Zotti}, {Delabrouille}, {D{\'e}mocl{\`e}s},
  {Diego}, {Dolag}, {Dole}, {Donzelli}, {Dor{\'e}}, {D{\"o}rl}, {Douspis},
  {Dupac}, {En{\ss}lin}, {Finelli}, {Flores-Cacho}, {Forni}, {Frailis},
  {Frommert}, {Galeotta}, {Ganga}, {G{\'e}nova-Santos}, {Giard}, {Giardino},
  {Gonz{\'a}alez-Nuevo}, {Gregorio}, {Gruppuso}, {Hansen}, {Harrison},
  {Hern{\'a}ndez-Monteagudo}, {Herranz}, {Hildebrandt}, {Hivon}, {Holmes},
  {Hovest}, {Huffenberger}, {Hurier}, {Jaffe}, {Jaffe}, {Jasche}, {Jones},
  {Juvela}, {Keih{\'a}nen}, {Keskitalo}, {Khamitov}, {Kisner}, {Knoche},
  {Kunz}, {Kurki-Suonio}, {Lagache}, {L{\"a}hteenm{\"a}ki}, {Lamarre},
  {Lasenby}, {Lawrence}, {Le Jeune}, {Leonardi}, {Lilje}, {Linden-V{\o}rnle},
  {L{\'o}pez-Caniego}, {Mac{\'\i}as-P{\'e}rez}, {Maino}, {Mak}, {Mandolesi},
  {Maris}, {Marleau}, {Mart{\'\i}nez-Gonz{\'a}lez}, {Masi}, {Matarrese},
  {Mazzotta}, {Melchiorri}, {Melin}, {Mendes}, {Mennella}, {Migliaccio},
  {Mitra}, {Miville-Desch{\^e}nes}, {Moneti}, {Montier}, {Morgante},
  {Mortlock}, {Moss}, {Munshi}, {Murphy}, {Naselsky}, {Nati}, {Natoli},
  {Netterfield}, {N{\o}rgaard-Nielsen}, {Noviello}, {Novikov}, {Novikov},
  {Osborne}, {Pagano}, {Paoletti}, {Perdereau}, {Perrotta}, {Piacentini},
  {Piat}, {Pierpaoli}, {Pietrobon}, {Plaszczynski}, {Pointecouteau}, {Polenta},
  {Popa}, {Poutanen}, {Pratt}, {Prunet}, {Puget}, {Puisieux}, {Rachen},
  {Rebolo}, {Reinecke}, {Remazeilles}, {Renault}, {Ricciardi}, {Roman},
  {Rubi{\~n}o-Mart{\'\i}n}, {Rusholme}, {Sandri}, {Savini}, {Scott}, {Spencer},
  {Sunyaev}, {Sutton}, {Suur-Uski}, {Sygnet}, {Tauber}, {Terenzi},
  {Toffolatti}, {Tomasi}, {Tristram}, {Tucci}, {Valenziano}, {Valiviita}, {Van
  Tent}, {Vielva}, {Villa}, {Vittorio}, {Wade}, {Welikala}, {Yvon}, {Zacchei},
  {Zibin}, \& {Zonca}}]{2014A&A...561A..97P}
{Planck Collaboration}, {Ade}, P.~A.~R., {Aghanim}, N., {et~al.} 2014, \aap,
  561, A97, \dodoi{10.1051/0004-6361/201321299}

\bibitem[{{Planck Collaboration} {et~al.}(2016){Planck Collaboration}, {Ade},
  {Aghanim}, {Arnaud}, {Ashdown}, {Aubourg}, {Aumont}, {Baccigalupi}, {Banday},
  {Barreiro}, {Bartolo}, {Battaner}, {Benabed}, {Benoit-L{\'e}vy},
  {Bersanelli}, {Bielewicz}, {Bock}, {Bonaldi}, {Bonavera}, {Bond}, {Borrill},
  {Bouchet}, {Burigana}, {Calabrese}, {Cardoso}, {Catalano}, {Chamballu},
  {Chiang}, {Christensen}, {Clements}, {Colombo}, {Combet}, {Crill}, {Curto},
  {Cuttaia}, {Danese}, {Davies}, {Davis}, {de Bernardis}, {de Zotti},
  {Delabrouille}, {Dickinson}, {Diego}, {Dolag}, {Donzelli}, {Dor{\'e}},
  {Douspis}, {Ducout}, {Dupac}, {Efstathiou}, {Elsner}, {En{\ss}lin},
  {Eriksen}, {Finelli}, {Forni}, {Frailis}, {Fraisse}, {Franceschi}, {Frejsel},
  {Galeotta}, {Galli}, {Ganga}, {G{\'e}nova-Santos}, {Giard}, {Gjerl{\o}w},
  {Gonz{\'a}lez-Nuevo}, {G{\'o}rski}, {Gregorio}, {Gruppuso}, {Hansen},
  {Harrison}, {Henrot-Versill{\'e}}, {Hern{\'a}ndez-Monteagudo}, {Herranz},
  {Hildebrandt}, {Hivon}, {Hobson}, {Hornstrup}, {Huffenberger}, {Hurier},
  {Jaffe}, {Jaffe}, {Jones}, {Juvela}, {Keih{\"a}nen}, {Keskitalo}, {Kitaura},
  {Kneissl}, {Knoche}, {Kunz}, {Kurki-Suonio}, {Lagache}, {Lamarre}, {Lasenby},
  {Lattanzi}, {Lawrence}, {Leonardi}, {Le{\'o}n-Tavares}, {Levrier}, {Liguori},
  {Lilje}, {Linden-V{\o}rnle}, {L{\'o}pez-Caniego}, {Lubin}, {Ma},
  {Mac{\'\i}as-P{\'e}rez}, {Maffei}, {Maino}, {Mak}, {Mandolesi}, {Mangilli},
  {Maris}, {Martin}, {Mart{\'\i}nez-Gonz{\'a}lez}, {Masi}, {Matarrese},
  {McGehee}, {Melchiorri}, {Mennella}, {Migliaccio}, {Miville-Desch{\^e}nes},
  {Moneti}, {Montier}, {Morgante}, {Mortlock}, {Munshi}, {Murphy}, {Naselsky},
  {Nati}, {Natoli}, {Noviello}, {Novikov}, {Novikov}, {Oxborrow}, {Pagano},
  {Pajot}, {Paoletti}, {Perdereau}, {Perotto}, {Pettorino}, {Piacentini},
  {Piat}, {Pierpaoli}, {Pointecouteau}, {Polenta}, {Ponthieu}, {Pratt},
  {Puget}, {Puisieux}, {Rachen}, {Racine}, {Reach}, {Reinecke}, {Remazeilles},
  {Renault}, {Renzi}, {Ristorcelli}, {Rocha}, {Rosset}, {Rossetti}, {Roudier},
  {Rubi{\~n}o-Mart{\'\i}n}, {Rusholme}, {Sandri}, {Santos}, {Savelainen},
  {Savini}, {Scott}, {Spencer}, {Stolyarov}, {Sudiwala}, {Sunyaev}, {Sutton},
  {Suur-Uski}, {Sygnet}, {Tauber}, {Terenzi}, {Toffolatti}, {Tomasi}, {Tucci},
  {Valenziano}, {Valiviita}, {Van Tent}, {Vielva}, {Villa}, {Wade}, {Wandelt},
  {Wang}, {Wehus}, {Yvon}, {Zacchei}, \& {Zonca}}]{2016A&A...586A.140P}
---. 2016, \aap, 586, A140, \dodoi{10.1051/0004-6361/201526328}

\bibitem[{{Planck Collaboration} {et~al.}(2018){Planck Collaboration},
  {Aghanim}, {Akrami}, {Ashdown}, {Aumont}, {Baccigalupi}, {Ballardini},
  {Banday}, {Barreiro}, {Bartolo}, {Basak}, {Battye}, {Benabed}, {Bernard},
  {Bersanelli}, {Bielewicz}, {Bond}, {Borrill}, {Bouchet}, {Burigana},
  {Calabrese}, {Carron}, {Chiang}, {Comis}, {Contreras}, {Crill}, {Curto},
  {Cuttaia}, {de Bernardis}, {de Rosa}, {de Zotti}, {Delabrouille}, {Di
  Valentino}, {Dickinson}, {Diego}, {Dor{\'e}}, {Ducout}, {Dupac}, {Elsner},
  {En{\ss}lin}, {Eriksen}, {Falgarone}, {Fantaye}, {Finelli}, {Forastieri},
  {Frailis}, {Fraisse}, {Franceschi}, {Frolov}, {Galeotta}, {Galli}, {Ganga},
  {Gerbino}, {G{\'o}rski}, {Gruppuso}, {Gudmundsson}, {Handley}, {Hansen},
  {Herranz}, {Hivon}, {Huang}, {Jaffe}, {Keih{\"a}nen}, {Keskitalo}, {Kiiveri},
  {Kim}, {Kisner}, {Krachmalnicoff}, {Kunz}, {Kurki-Suonio}, {Lamarre},
  {Lasenby}, {Lattanzi}, {Lawrence}, {Le Jeune}, {Levrier}, {Liguori}, {Lilje},
  {Lindholm}, {L{\'o}pez-Caniego}, {Lubin}, {Ma}, {Mac{\'\i}as-P{\'e}rez},
  {Maggio}, {Maino}, {Mandolesi}, {Mangilli}, {Martin},
  {Mart{\'\i}nez-Gonz{\'a}lez}, {Matarrese}, {Mauri}, {McEwen}, {Melchiorri},
  {Mennella}, {Migliaccio}, {Miville-Desch{\^e}nes}, {Molinari}, {Moneti},
  {Montier}, {Morgante}, {Natoli}, {Oxborrow}, {Pagano}, {Paoletti},
  {Partridge}, {Perdereau}, {Perotto}, {Pettorino}, {Piacentini},
  {Plaszczynski}, {Polastri}, {Polenta}, {Rachen}, {Racine}, {Reinecke},
  {Remazeilles}, {Renzi}, {Rocha}, {Roudier}, {Ruiz-Granados}, {Sandri},
  {Savelainen}, {Scott}, {Sirignano}, {Sirri}, {Spencer}, {Stanco}, {Sunyaev},
  {Tauber}, {Tavagnacco}, {Tenti}, {Toffolatti}, {Tomasi}, {Tristram},
  {Trombetti}, {Valiviita}, {Van Tent}, {Vielva}, {Villa}, {Vittorio},
  {Wandelt}, {Wehus}, {Zacchei}, \& {Zonca}}]{2018A&A...617A..48P}
{Planck Collaboration}, {Aghanim}, N., {Akrami}, Y., {et~al.} 2018, \aap, 617,
  A48, \dodoi{10.1051/0004-6361/201731489}

\bibitem[{{Reichardt} {et~al.}(2021){Reichardt}, {Patil}, {Ade}, {Anderson},
  {Austermann}, {Avva}, {Baxter}, {Beall}, {Bender}, {Benson}, {Bianchini},
  {Bleem}, {Carlstrom}, {Chang}, {Chaubal}, {Chiang}, {Chou}, {Citron},
  {Moran}, {Crawford}, {Crites}, {de Haan}, {Dobbs}, {Everett}, {Gallicchio},
  {George}, {Gilbert}, {Gupta}, {Halverson}, {Harrington}, {Henning}, {Hilton},
  {Holder}, {Holzapfel}, {Hrubes}, {Huang}, {Hubmayr}, {Irwin}, {Knox}, {Lee},
  {Li}, {Lowitz}, {Luong-Van}, {McMahon}, {Mehl}, {Meyer}, {Millea}, {Mocanu},
  {Mohr}, {Montgomery}, {Nadolski}, {Natoli}, {Nibarger}, {Noble}, {Novosad},
  {Omori}, {Padin}, {Pryke}, {Ruhl}, {Saliwanchik}, {Sayre}, {Schaffer},
  {Shirokoff}, {Sievers}, {Smecher}, {Spieler}, {Staniszewski}, {Stark},
  {Tucker}, {Vanderlinde}, {Veach}, {Vieira}, {Wang}, {Whitehorn},
  {Williamson}, {Wu}, \& {Yefremenko}}]{2021ApJ...908..199R}
{Reichardt}, C.~L., {Patil}, S., {Ade}, P.~A.~R., {et~al.} 2021, \apj, 908,
  199, \dodoi{10.3847/1538-4357/abd407}

\bibitem[{{Santos} {et~al.}(2003){Santos}, {Cooray}, {Haiman}, {Knox}, \&
  {Ma}}]{2003ApJ...598..756S}
{Santos}, M.~G., {Cooray}, A., {Haiman}, Z., {Knox}, L., \& {Ma}, C.-P. 2003,
  \apj, 598, 756, \dodoi{10.1086/378772}

\bibitem[{{Sayers} {et~al.}(2013){Sayers}, {Mroczkowski}, {Zemcov}, {Korngut},
  {Bock}, {Bulbul}, {Czakon}, {Egami}, {Golwala}, {Koch}, {Lin}, {Mantz},
  {Molnar}, {Moustakas}, {Pierpaoli}, {Rawle}, {Reese}, {Rex}, {Shitanishi},
  {Siegel}, \& {Umetsu}}]{2013ApJ...778...52S}
{Sayers}, J., {Mroczkowski}, T., {Zemcov}, M., {et~al.} 2013, \apj, 778, 52,
  \dodoi{10.1088/0004-637X/778/1/52}

\bibitem[{{Sayers} {et~al.}(2016){Sayers}, {Zemcov}, {Glenn}, {Golwala},
  {Maloney}, {Siegel}, {Wheeler}, {Bockstiegel}, {Brugger}, {Czakon}, {Day},
  {Downes}, {Duan}, {Gao}, {Hollister}, {Lam}, {LeDuc}, {Mazin}, {McHugh},
  {Miller}, {Mroczkowski}, {Noroozian}, {Nguyen}, {Radford}, {Schlaerth},
  {Vayonakis}, {Wilson}, \& {Zmuidzinas}}]{2016ApJ...820..101S}
{Sayers}, J., {Zemcov}, M., {Glenn}, J., {et~al.} 2016, \apj, 820, 101,
  \dodoi{10.3847/0004-637X/820/2/101}

\bibitem[{{Schaan} {et~al.}(2016){Schaan}, {Ferraro}, {Vargas-Maga{\~n}a},
  {Smith}, {Ho}, {Aiola}, {Battaglia}, {Bond}, {De Bernardis}, {Calabrese},
  {Cho}, {Devlin}, {Dunkley}, {Gallardo}, {Hasselfield}, {Henderson}, {Hill},
  {Hincks}, {Hlozek}, {Hubmayr}, {Hughes}, {Irwin}, {Koopman}, {Kosowsky},
  {Li}, {Louis}, {Lungu}, {Madhavacheril}, {Maurin}, {McMahon}, {Moodley},
  {Naess}, {Nati}, {Newburgh}, {Niemack}, {Page}, {Pappas}, {Partridge},
  {Schmitt}, {Sehgal}, {Sherwin}, {Sievers}, {Spergel}, {Staggs}, {van
  Engelen}, {Wollack}, \& {ACTPol Collaboration}}]{2016PhRvD..93h2002S}
{Schaan}, E., {Ferraro}, S., {Vargas-Maga{\~n}a}, M., {et~al.} 2016, \prd, 93,
  082002, \dodoi{10.1103/PhysRevD.93.082002}

\bibitem[{{Schaan} {et~al.}(2021){Schaan}, {Ferraro}, {Amodeo}, {Battaglia},
  {Aiola}, {Austermann}, {Beall}, {Bean}, {Becker}, {Bond}, {Calabrese},
  {Calafut}, {Choi}, {Denison}, {Devlin}, {Duff}, {Duivenvoorden}, {Dunkley},
  {D{\"u}nner}, {Gallardo}, {Guan}, {Han}, {Hill}, {Hilton}, {Hilton},
  {Hlo{\v{z}}ek}, {Hubmayr}, {Huffenberger}, {Hughes}, {Koopman}, {MacInnis},
  {McMahon}, {Madhavacheril}, {Moodley}, {Mroczkowski}, {Naess}, {Nati},
  {Newburgh}, {Niemack}, {Page}, {Partridge}, {Salatino}, {Sehgal},
  {Schillaci}, {Sif{\'o}n}, {Smith}, {Spergel}, {Staggs}, {Storer}, {Trac},
  {Ullom}, {Van Lanen}, {Vale}, {van Engelen}, {Maga{\~n}a}, {Vavagiakis},
  {Wollack}, {Xu}, \& {Atacama Cosmology Telescope
  Collaboration}}]{2021PhRvD.103f3513S}
{Schaan}, E., {Ferraro}, S., {Amodeo}, S., {et~al.} 2021, \prd, 103, 063513,
  \dodoi{10.1103/PhysRevD.103.063513}

\bibitem[{{Shao} \& {Fang}(2016)}]{2016MNRAS.458.3773S}
{Shao}, J., \& {Fang}, T. 2016, \mnras, 458, 3773, \dodoi{10.1093/mnras/stw501}

\bibitem[{{Shaw} {et~al.}(2012){Shaw}, {Rudd}, \&
  {Nagai}}]{2012ApJ...756...15S}
{Shaw}, L.~D., {Rudd}, D.~H., \& {Nagai}, D. 2012, \apj, 756, 15,
  \dodoi{10.1088/0004-637X/756/1/15}

\bibitem[{{Soergel} {et~al.}(2018){Soergel}, {Saro}, {Giannantonio},
  {Efstathiou}, \& {Dolag}}]{2018MNRAS.478.5320S}
{Soergel}, B., {Saro}, A., {Giannantonio}, T., {Efstathiou}, G., \& {Dolag}, K.
  2018, \mnras, 478, 5320, \dodoi{10.1093/mnras/sty1324}

\bibitem[{{Soergel} {et~al.}(2016){Soergel}, {Flender}, {Story}, {Bleem},
  {Giannantonio}, {Efstathiou}, {Rykoff}, {Benson}, {Crawford}, {Dodelson},
  {Habib}, {Heitmann}, {Holder}, {Jain}, {Rozo}, {Saro}, {Weller}, {Abdalla},
  {Allam}, {Annis}, {Armstrong}, {Benoit-L{\'e}vy}, {Bernstein}, {Carlstrom},
  {Carnero Rosell}, {Carrasco Kind}, {Castander}, {Chiu}, {Chown}, {Crocce},
  {Cunha}, {D'Andrea}, {da Costa}, {de Haan}, {Desai}, {Diehl}, {Dietrich},
  {Doel}, {Estrada}, {Evrard}, {Flaugher}, {Fosalba}, {Frieman}, {Gaztanaga},
  {Gruen}, {Gruendl}, {Holzapfel}, {Honscheid}, {James}, {Keisler}, {Kuehn},
  {Kuropatkin}, {Lahav}, {Lima}, {Marshall}, {McDonald}, {Melchior}, {Miller},
  {Miquel}, {Nord}, {Ogando}, {Omori}, {Plazas}, {Rapetti}, {Reichardt},
  {Romer}, {Roodman}, {Saliwanchik}, {Sanchez}, {Schubnell}, {Sevilla-Noarbe},
  {Sheldon}, {Smith}, {Soares-Santos}, {Sobreira}, {Stark}, {Suchyta},
  {Swanson}, {Tarle}, {Thomas}, {Vieira}, {Walker}, {Whitehorn}, {DES
  Collaboration}, \& {SPT Collaboration}}]{2016MNRAS.461.3172S}
{Soergel}, B., {Flender}, S., {Story}, K.~T., {et~al.} 2016, \mnras, 461, 3172,
  \dodoi{10.1093/mnras/stw1455}

\bibitem[{{Sugiyama} {et~al.}(2017){Sugiyama}, {Okumura}, \&
  {Spergel}}]{2017JCAP...01..057S}
{Sugiyama}, N.~S., {Okumura}, T., \& {Spergel}, D.~N. 2017, \jcap, 2017, 057,
  \dodoi{10.1088/1475-7516/2017/01/057}

\bibitem[{{Sugiyama} {et~al.}(2018){Sugiyama}, {Okumura}, \&
  {Spergel}}]{2018MNRAS.475.3764S}
---. 2018, \mnras, 475, 3764, \dodoi{10.1093/mnras/stx3362}

\bibitem[{{Sunyaev} \& {Zeldovich}(1972)}]{SZ72}
{Sunyaev}, R.~A., \& {Zeldovich}, Y.~B. 1972, Comments on Astrophysics and
  Space Physics, 4, 173

\bibitem[{{Sunyaev} \& {Zeldovich}(1980)}]{1980MNRAS.190..413S}
---. 1980, \mnras, 190, 413, \dodoi{10.1093/mnras/190.3.413}

\bibitem[{{Tanimura} {et~al.}(2021){Tanimura}, {Zaroubi}, \&
  {Aghanim}}]{2021A&A...645A.112T}
{Tanimura}, H., {Zaroubi}, S., \& {Aghanim}, N. 2021, \aap, 645, A112,
  \dodoi{10.1051/0004-6361/202038846}

\bibitem[{{Terrana} {et~al.}(2017){Terrana}, {Harris}, \&
  {Johnson}}]{2017JCAP...02..040T}
{Terrana}, A., {Harris}, M.-J., \& {Johnson}, M.~C. 2017, \jcap, 2017, 040,
  \dodoi{10.1088/1475-7516/2017/02/040}

\bibitem[{{Tokutake} {et~al.}(2018){Tokutake}, {Ichiki}, \&
  {Yoo}}]{2018JCAP...03..033T}
{Tokutake}, M., {Ichiki}, K., \& {Yoo}, C.-M. 2018, \jcap, 2018, 033,
  \dodoi{10.1088/1475-7516/2018/03/033}

\bibitem[{{Vavagiakis} {et~al.}(2021){Vavagiakis}, {Gallardo}, {Calafut},
  {Amodeo}, {Aiola}, {Austermann}, {Battaglia}, {Battistelli}, {Beall}, {Bean},
  {Bond}, {Calabrese}, {Choi}, {Cothard}, {Devlin}, {Duell}, {Duivenvoorden},
  {Dunkley}, {Dunner}, {Ferraro}, {Guan}, {Hill}, {Hilton}, {Hlozek}, {Huber},
  {Hubmayr}, {Huffenberger}, {Hughes}, {Koopman}, {Kosowsky}, {Li}, {Lokken},
  {Madhavacheril}, {McMahon}, {Moodley}, {Naess}, {Nati}, {Newburgh},
  {Niemack}, {Partridge}, {Schaan}, {Schillaci}, {Sifon}, {Spergel}, {Staggs},
  {Ullom}, {Vale}, {Van Engelen}, {Wollack}, \& {Xu}}]{2021arXiv210108373V}
{Vavagiakis}, E.~M., {Gallardo}, P.~A., {Calafut}, V., {et~al.} 2021, arXiv
  e-prints, arXiv:2101.08373.
\newblock \doarXiv{2101.08373}

\bibitem[{{Wang} {et~al.}(2020){Wang}, {Ramachandra}, {Salazar-Canizales},
  {Feldman}, {Watkins}, \& {Dolag}}]{2020arXiv201003762W}
{Wang}, Y., {Ramachandra}, N., {Salazar-Canizales}, E.~M., {et~al.} 2020, arXiv
  e-prints, arXiv:2010.03762.
\newblock \doarXiv{2010.03762}

\bibitem[{{Xu} {et~al.}(2015){Xu}, {Wang}, \& {Zhang}}]{2015PhRvD..92h3505X}
{Xu}, X.-d., {Wang}, B., \& {Zhang}, P. 2015, \prd, 92, 083505,
  \dodoi{10.1103/PhysRevD.92.083505}

\bibitem[{{Xu} {et~al.}(2013){Xu}, {Wang}, {Zhang}, \&
  {Atrio-Barandela}}]{2013JCAP...12..001X}
{Xu}, X.-D., {Wang}, B., {Zhang}, P., \& {Atrio-Barandela}, F. 2013, \jcap,
  2013, 001, \dodoi{10.1088/1475-7516/2013/12/001}

\bibitem[{{Yang} {et~al.}(2005){Yang}, {Mo}, {van den Bosch}, \&
  {Jing}}]{2005MNRAS.356.1293Y}
{Yang}, X., {Mo}, H.~J., {van den Bosch}, F.~C., \& {Jing}, Y.~P. 2005, \mnras,
  356, 1293, \dodoi{10.1111/j.1365-2966.2005.08560.x}

\bibitem[{Yang {et~al.}(2020)Yang, Xu, He, Gu, Katsianis, Meng, Shi, Zou,
  Zhang, Liu, Wang, Dong, Lu, Li, Chen, Wang, Mo, Fu, Guo, Leauthaud, Luo,
  Zhang, \& Zu}]{yang2020extended}
Yang, X., Xu, H., He, M., {et~al.} 2020, An Extended Halo-based Group/Cluster
  finder: application to the DESI legacy imaging surveys DR8.
\newblock \doarXiv{2012.14998}

\bibitem[{{Zeldovich} \& {Sunyaev}(1969)}]{SZ69}
{Zeldovich}, Y.~B., \& {Sunyaev}, R.~A. 1969, \apss, 4, 301,
  \dodoi{10.1007/BF00661821}

\bibitem[{{Zhang}(2010)}]{2010MNRAS.407L..36Z}
{Zhang}, P. 2010, \mnras, 407, L36, \dodoi{10.1111/j.1745-3933.2010.00899.x}

\bibitem[{{Zhang} {et~al.}(2008){Zhang}, {Feldman}, {Juszkiewicz}, \&
  {Stebbins}}]{2008MNRAS.388..884Z}
{Zhang}, P., {Feldman}, H.~A., {Juszkiewicz}, R., \& {Stebbins}, A. 2008,
  \mnras, 388, 884, \dodoi{10.1111/j.1365-2966.2008.13454.x}

\bibitem[{{Zhang} \& {Johnson}(2015)}]{2015JCAP...06..046Z}
{Zhang}, P., \& {Johnson}, M.~C. 2015, \jcap, 2015, 046,
  \dodoi{10.1088/1475-7516/2015/06/046}

\bibitem[{{Zhang} {et~al.}(2004){Zhang}, {Pen}, \&
  {Trac}}]{2004MNRAS.347.1224Z}
{Zhang}, P., {Pen}, U.-L., \& {Trac}, H. 2004, \mnras, 347, 1224,
  \dodoi{10.1111/j.1365-2966.2004.07298.x}

\bibitem[{{Zhang} \& {Stebbins}(2011)}]{2011PhRvL.107d1301Z}
{Zhang}, P., \& {Stebbins}, A. 2011, \prl, 107, 041301,
  \dodoi{10.1103/PhysRevLett.107.041301}

\bibitem[{{Zheng}(2020)}]{2020ApJ...904...48Z}
{Zheng}, Y. 2020, \apj, 904, 48, \dodoi{10.3847/1538-4357/abbb99}

\end{thebibliography}
\bibliographystyle{aasjournal}

\appendix
\section{Pairwise velocity of halos in simulation and the fitting template}
 \label{app:simulation}
	\begin{figure*}
	\centering
	\includegraphics[width=1\textwidth]{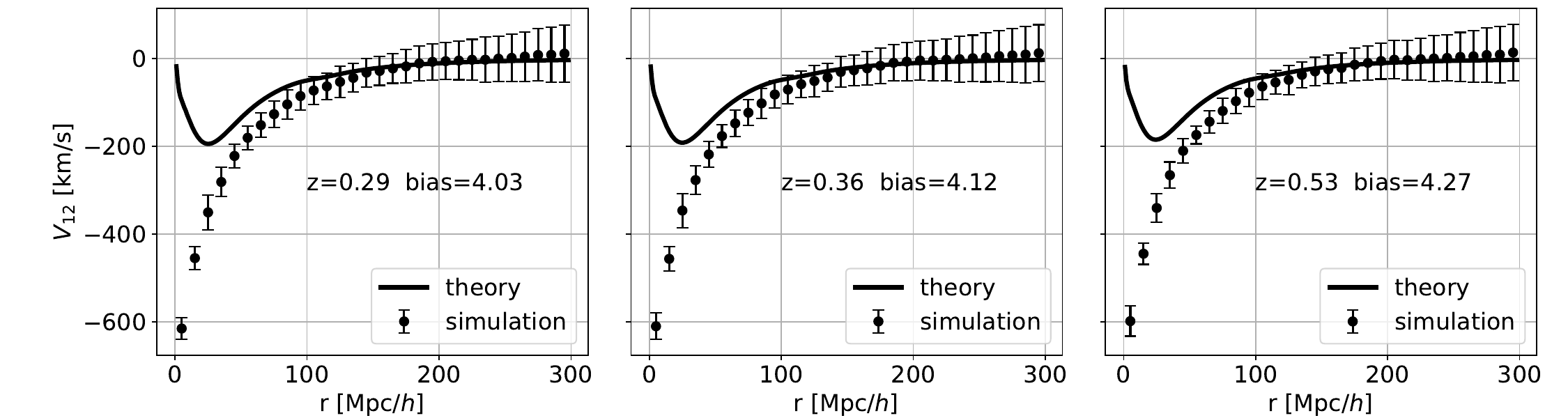}
 	\caption{The simulated pairwise velocity (data points with error bars) and a  linear theory prediction (Eq. \ref{eq:pairwise_velocity_theory}solid curves). The linear theory predictions show significant inaccuracy at $\lesssim 50\ {\rm Mpc}/h$. Although photo-z errors reduces the significance of this inaccuracy, we will use the numerically simulated $v_{12}$ as the template of interpreting the kSZ measurement. Notice that the simulated $v_{12}$ drops to zero at $r\ll 5\ {\rm Mpc}/h$, but this behavior does not show up here due to coarse bin size. \label{fig:app_simu_v12}}
	\end{figure*}
	 	\begin{figure*}
	\centering
	\includegraphics[width=1\textwidth]{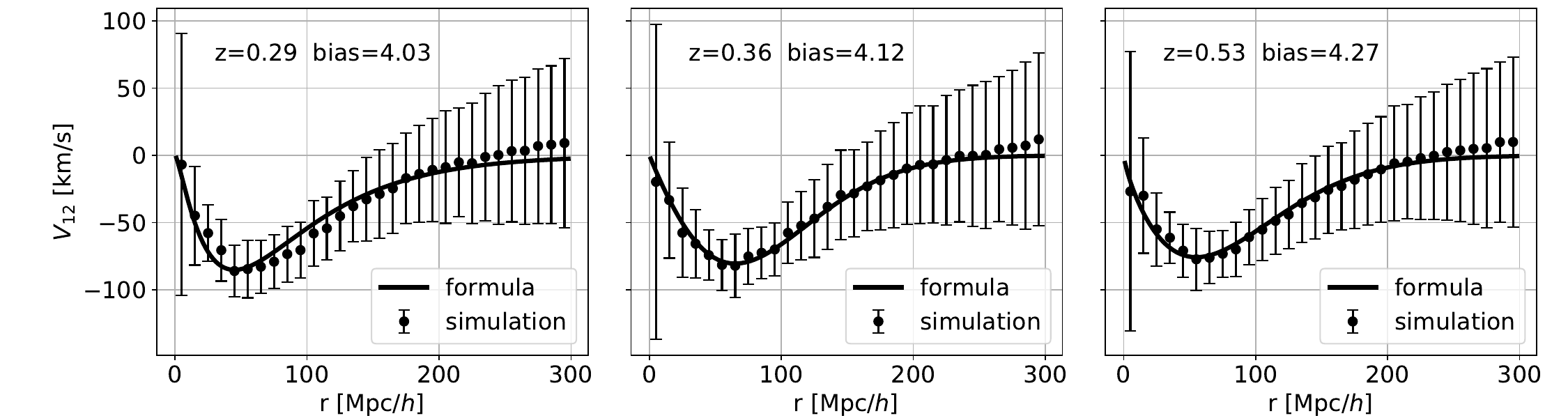}
 	\caption{The simulated pairwise velocity of halos with photo-z errors (points with errorbars). The solid lines are the fit with $v_{12}(r)=v_{\rm p}(r/r_{\rm p})^\alpha e^{-(\alpha/\beta)(1-(r/r_{\rm p})^\beta)}$.\label{fig:app_simu_v12_zph}}
	\end{figure*}
	
 	\begin{figure}
 	\centering
	\includegraphics[width=0.5\textwidth]{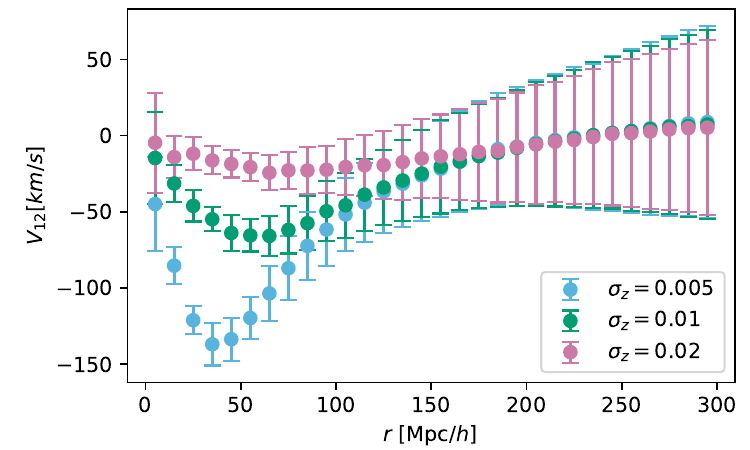}
 	\caption{The pairwise velocity of a halo sample whose bias is 3.11 at z=0.326 with different photo-z error 0.005, 0.01 and 0.02 (1+z). \label{fig:app_simu_influence_zph}}
	\end{figure}
	
	\begin{figure}
 	\centering
	\includegraphics[width=0.45\textwidth]{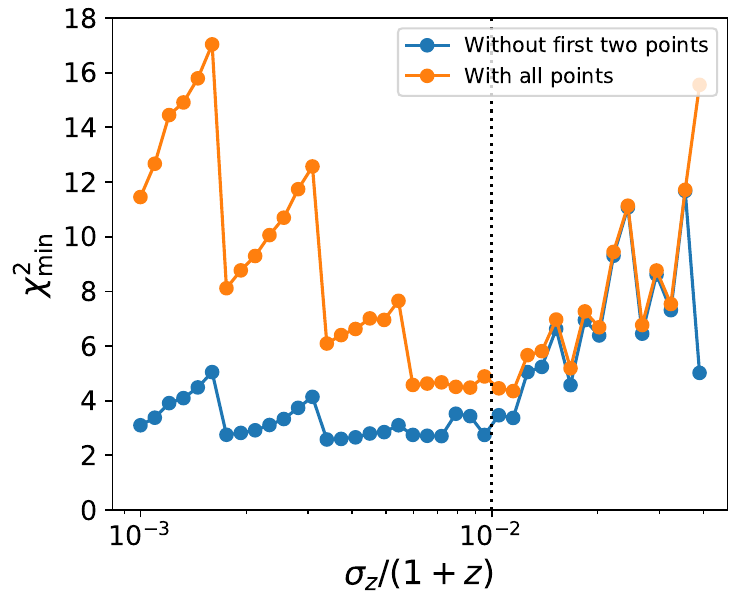}
 	\caption{$\chi^2_{\rm min}$ as a function of $\sigma_z/(1+z)$. Blue line excludes the first two measurement points as in the main body. Orange line utilize all measurement points, and it reaches the minimum value at $\sigma_z/(1+z) \sim 0.01$.\label{fig:chi2_sigmaz}}
	\end{figure}

The pairwise velocity $v_{12}$ depends on both the halo mass and redshift. As examples, we measure $v_{12}(r)$ at  three halo samples of different redshift and bias (Fig. \ref{fig:app_simu_v12}), using one of the CosmicGrowth simulations.  To estimate the simulation uncertainty, we divide the simulation box into 8 sub-boxes (volume = $600^3 [{\rm Mpc}/h]^3$). Since there are 3 independent directions, we have 24 independent samples to estimate the error bars (Fig. \ref{fig:app_simu_v12}).  We also compare the numerically obtained $v_{12}$ to a widely adopted theoretical template \citep{2010gfe..book.....M, Mueller_2015}, 
\ba
	\label{eq:pairwise_velocity_theory}
	v_{12}(r, z)=-\frac{2}{3} H a f \frac{r \bar{\xi}_{\rm c}(r,z)}{1+\xi_{\rm c}(r, z)}\ .
\ea	
Here $\xi_{\rm c}(r, z)$ is the 2-points correlation function of galaxy cluster. And $ \bar{\xi}_{\rm c}(r,z)$ is the volume averaged correlation function. $H$ is the Hubble parameter and $f\equiv d\ln D/d\ln a$ is the logarithmic growth rate. The above template becomes inaccurate at $r < 50 {\rm\  Mpc}/h$,  showing the failure of linear theory.

Next we add ramdom shift into the simulated halos to mimic observed galaxy clusters with photo-z errors.  We adopt a Gaussian photo-z error PDF.  $|v_{12}|$ decreases significantly at $r\la c\sigma_z/H$ (Fig. \ref{fig:app_simu_v12_zph}). It peaks at $r\sim 60\ {\rm Mpc}/h$ and the peak position moves to larger $r$ with increasing redshift. $v_{12}$ should approach zero monotonically when $r\rightarrow \infty$. However, due to statistical fluctuations, the numerically measured $v_{12}$ scatters from zero at large $r$. To reduce such numerical fluctuations, we fit the simulated result with $v_{12}(r)=v_{\rm p}(r/r_{\rm p})^\alpha e^{-(\alpha/\beta)(1-(r/r_{\rm p})^\beta)}$. Here $v_{\rm p}$  and $r_{\rm p}$ are the peak amplitude and peak position respectively. $\alpha$ and $\beta$ are the other two free parameters to fit.  This fitting formula describes the simulation excellently (Fig. \ref{fig:app_simu_v12_zph}), and we will use it as the template for interpreting the pairwise kSZ measurement. For different cluster samples, we use different template from simulated halos, with matching mean redshift and mean bias. 

	We also test how the photo-z error would influence the pairwise velocity. We choose $\sigma_z/(1+z)=0.005,\ 0.01,\ 0.02$ and the results are shown in Fig. \ref{fig:app_simu_influence_zph}. With the increasing of photo-z error, we find the amplitude of pairwise velocity is decreasing rapidly and the peak moves to larger $r$. For a large photo-z error ($\sigma_z/(1+z)=0.02$), the pairwise velocity is almost consistent with null signal. The parameter $\sigma_z$ is an unknown value of a specific cluster sample and it will influence the shape and amplitude of the template a lot. In the measurement, we use the number $\sigma_z/(1+z)=0.01$ given by (\cite{yang2020extended}). 
	To further test whether $\sigma_z/(1+z)$ equals $0.01$ for the baseline sample, we set $\sigma_z/(1+z)$ as a free parameter. Fig. \ref{fig:chi2_sigmaz} shows $\chi_{\rm min}$ as a function of $\sigma_z/(1+z)$. With all measurement points, $\chi_{\rm min}$ reaches the minimum value at $\sigma_z/(1+z) \sim 0.01$ as expected.

	\begin{figure}
	\centering
	\includegraphics[width=0.5
	\textwidth]{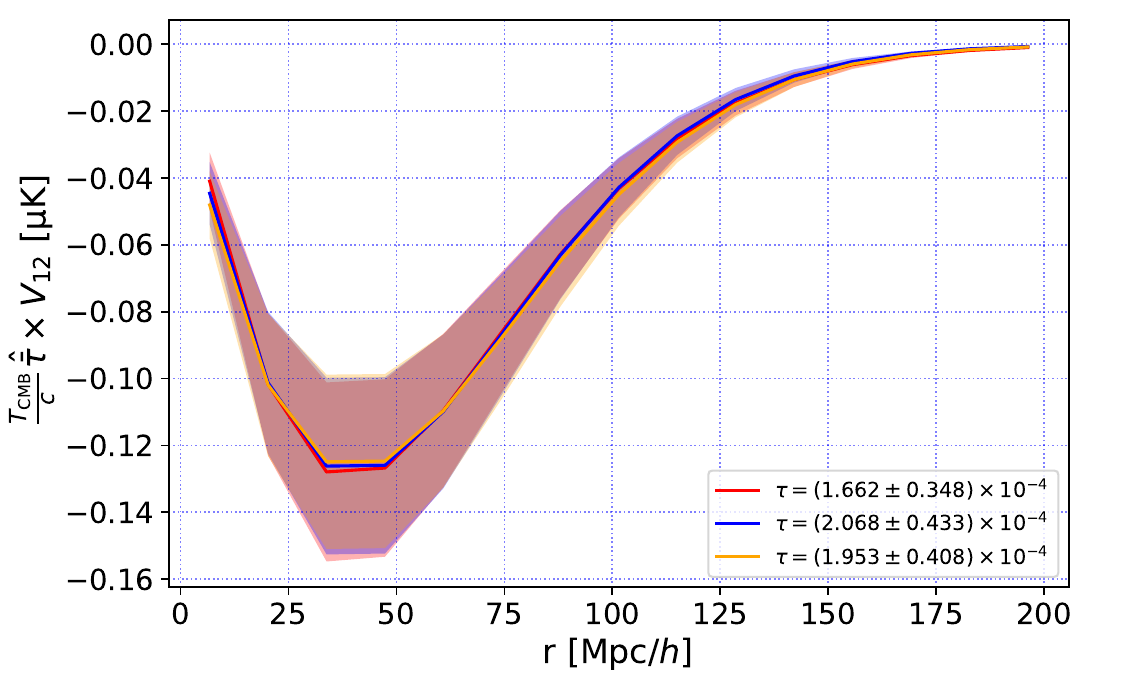}
 	\caption{The dependence of kSZ measurement on the adopted $v_{12}(r)$ template. Although the bestfit $\bar{\tau}_e$ varies with the template, both $\bar{\tau} \times v_{12}$ and the S/N ($\bar{\tau}_e/\sigma_\tau$ do not. Therefore for comparisons between different works, we should compare $\bar{\tau} \times v_{12}$ instead of $\bar{\tau}$ to avoid such uncertainty. This figure also shows that the inferred S/N is robust against the template uncertainty. 
 \label{fig:app_template_dependence} \label{fig:app_temp_depend}}
	\end{figure}

We caution that the constraint on the mean optical depth $\bar{\tau}_e$ is degenerate with the adopted template of $v_{12}$. The template for the baseline cluster sample has $\bar{z}=0.32$ and $\bar{b}_g=3.11$. If we use other templates, constraints on $\bar{\tau}_e$ will change. To demonstrate this point, we choose three other templates with  $z$ = 0.289, 0.528, 0.326 and $b$ = 4.03, 3.96, 3.44.  The results are shown in Fig. \ref{fig:app_template_dependence}. Indeed $\bar{\tau}_e$ varies with the template, while  the product $\hat {\bar{\tau}} \times v_{12}$ remains essentially unchanged.  The most important point that we find is that the S/N is essentially independent of the template adopted. The S/N for the three templates are $4.78$, $4.78$, $4.79$, versus $4.69$ of the baseline template. Therefore, despite uncertainties in the theoretical template, the S/N of the kSZ measurement is robust.

\section{Different estimators}\label{app:different_models}
	\begin{figure*}
	    \centering
    	\includegraphics[width=1\textwidth]{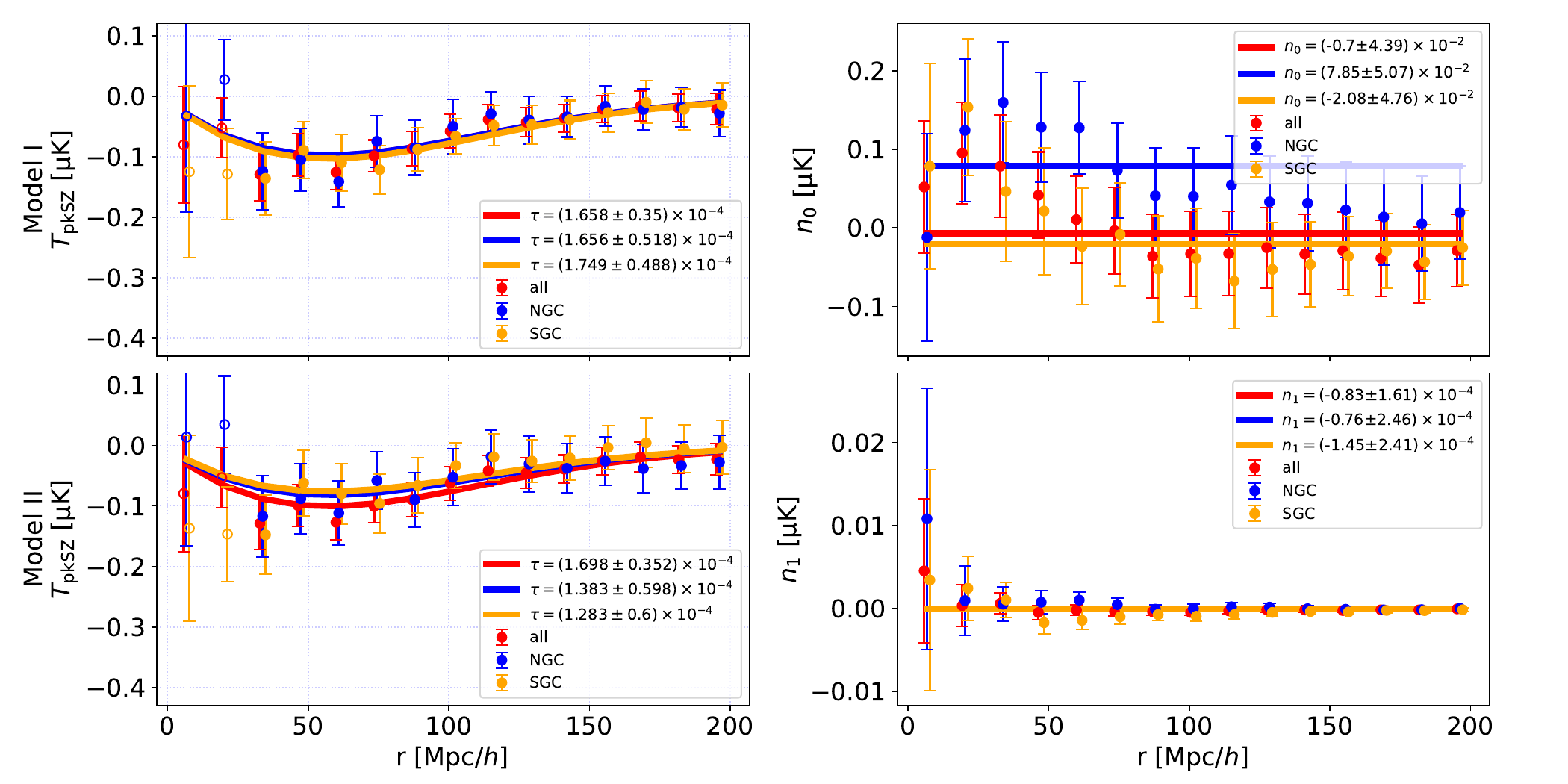}
	    \caption{The left two panels are the pairwise kSZ signal measured adopt Model I and Model II, respectively. The top-right panel is the $n_0$ term measured in Model I. The top-bottom panel is the $n_1$ term measured in Model II. Model II is found to be inappropriate for the kSZ measurement, since both $\tau$ and $n_1$ are inconsistent with the full model (baseline model). \label{fig:app_2_model}}
	\end{figure*}

The baseline model considers both systematic errors of $n_0$ and $n_1$. One question is whether it is necessary to include both of them. For this purpose, we consider two other models. Model I only includes $n_0$, 
	\ba
		\textbf{Model\ I}:\ 	T_{ij}^{\rm theory}=\hat{T}_{\rm pkSZ}C_{ij}+\hat n_{0}\ .
	\ea
The estimator is 
	\begin{eqnarray}
		\textbf{Model\ I}: \quad
		\hat T_{\rm pkSZ}&=&\frac{\langle TC \rangle -\langle T\rangle \langle C\rangle }{\langle C^2\rangle -\langle C\rangle ^2} \ , \\
		 \hat n_0 &=&\langle T\rangle -\hat T_{\rm pkSZ}\langle C\rangle \ .
	\end{eqnarray}
Model II only considers $n_1$, 
	\begin{eqnarray}\label{eq:model2}
	\textbf{Model\ II}:\ 		T_{\rm ij}^{\rm theory}&=&\hat{T}_{\rm pkSZ}C_{ij}+\hat n_1 z_{ij}\ ,\\ 
			z_{ij}&\equiv& z_i-z_j \ .
	\end{eqnarray}
The estimator is 
	\begin{eqnarray}
		\textbf{Model\ II}: \quad
		\hat T_{\rm pkSZ}&=&\frac{\langle TC\rangle \langle Z^2\rangle -\langle TZ\rangle \langle CZ\rangle }{\langle C^2Z^2\rangle -\langle Z^2\rangle \langle C^2\rangle}\ ,\\
		 \hat n_1&=&\frac{\langle TC\rangle \langle Z^2\rangle -\langle TZ\rangle \langle CZ\rangle }{\langle C^2Z^2\rangle -\langle Z^2\rangle \langle C^2\rangle }\ .
	\end{eqnarray}
The measured $T_{\rm pkSZ}$ and $n_{0,1}$ are shown in Fig. \ref{fig:app_2_model}. Comparing with the baseline model, we find that model I produces consistent $T_{\rm pkSZ}$, while model II does not. Furthermore, $n_1$ in model II differs from the baseline model significantly. And the noise term of each model is shown in right panels of Fig. \ref{fig:app_2_model}.  We then conclude that the $n_0$ type systematic error is present in the data and has to be included in the analysis. But the $n_1$ type is negligible. This is also consistent with the finding of tiny value of $n_1$ in the baseline model. 
	
    
    Compare with the baseline model in Section \ref{subsec:analysis_pairwise_kSZ_estimator}, we find that the $n_1$ term in the baseline model is much smaller than that of Model II, which means the $n_0$ and $n_1$ term are not independent. For both the baseline model and Model I, $n_0$ derives from zero more than 1$\sigma$ for NGC sample, which proves $n_0$ should be a component in the noise residual. And for Model II, in which the $n_0$ term is ignored, some difference is raised for three sample. Therefore, the conclusion can be made that only a redshift-dependent term is not suitable to describe the residual noise. The measurements for the baseline model and Model I are similar due to the negligible amplitude of $n_1$ comparing to $n_0$.

\section{Different CMB maps}\label{app:different_cmb_maps}

	\begin{figure*}
	\centering
	\includegraphics[width=1\textwidth]{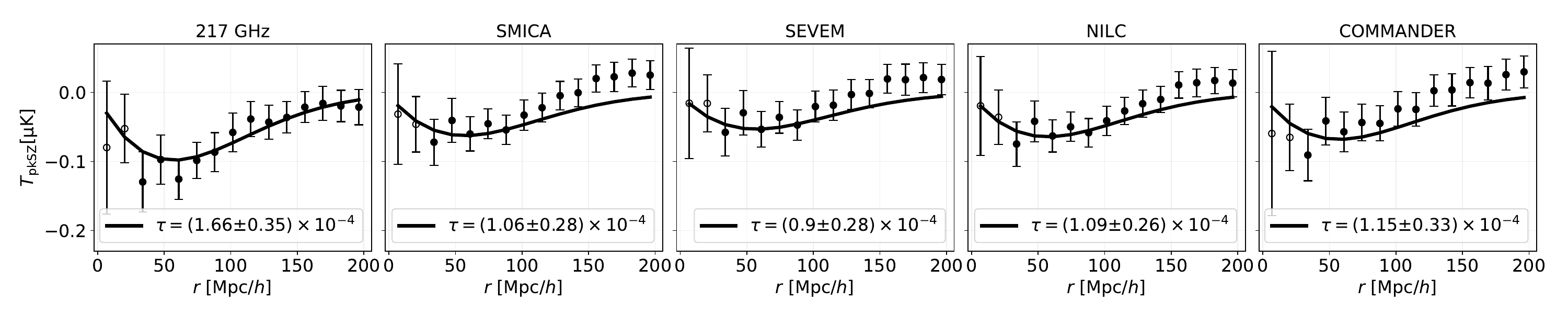}
	\caption{The pariwise kSZ measurements of five CMB maps.  The four foreground-cleaned maps have worse angular resolution, and therefore smaller $\bar{\tau}$.\label{fig:app_diff_cmb_pksz}}
	\end{figure*}

	\begin{figure*}
	\centering
	\includegraphics[width=1\textwidth]{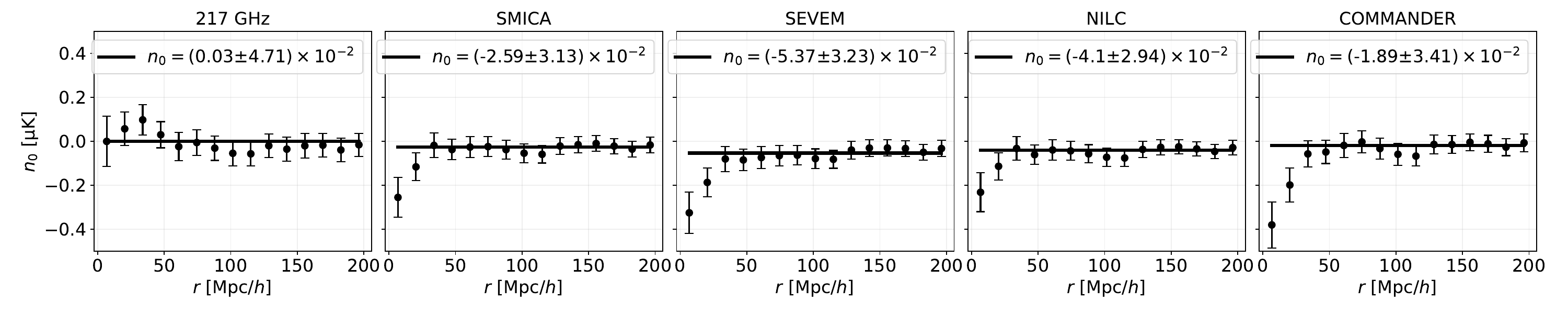}
	\caption{The noise term $n_0$ of five CMB maps. We detect $n_0$ for all four foreground-cleaned maps, which may be caused by residual thermal SZ effect in clusters. \label{fig:app_diff_cmb_noise}}
	\end{figure*}
	
	Planck survey provides several foreground-cleaned maps (SMICA, SEVEM, NLIC and COMMANDER). We compare the pairwise kSZ measurements of them with the 217 GHz result (Fig. \ref{fig:app_diff_cmb_pksz}). The amplitude of pairwise kSZ using these foreground-cleaned maps is systematically lower. This is caused by  larger (effective) beam of these maps which further dilutes the kSZ signal. But the major  difference is in $n_0$.  $n_0\neq 0$ is detected in all these 4 maps, especially for the first two $r$-bins. This may be due to residual tSZ effect of clusters in these foreground-cleaned maps. For the two reasons, we conclude that the 217 GHz map is the most optimal for the pairwise kSZ measurement.

\section{Covariance matrix and S/N estimation}\label{app:consistency_SN}

    \begin{figure*}
	\centering
	\includegraphics[width=0.9\textwidth, height=0.36\textwidth]{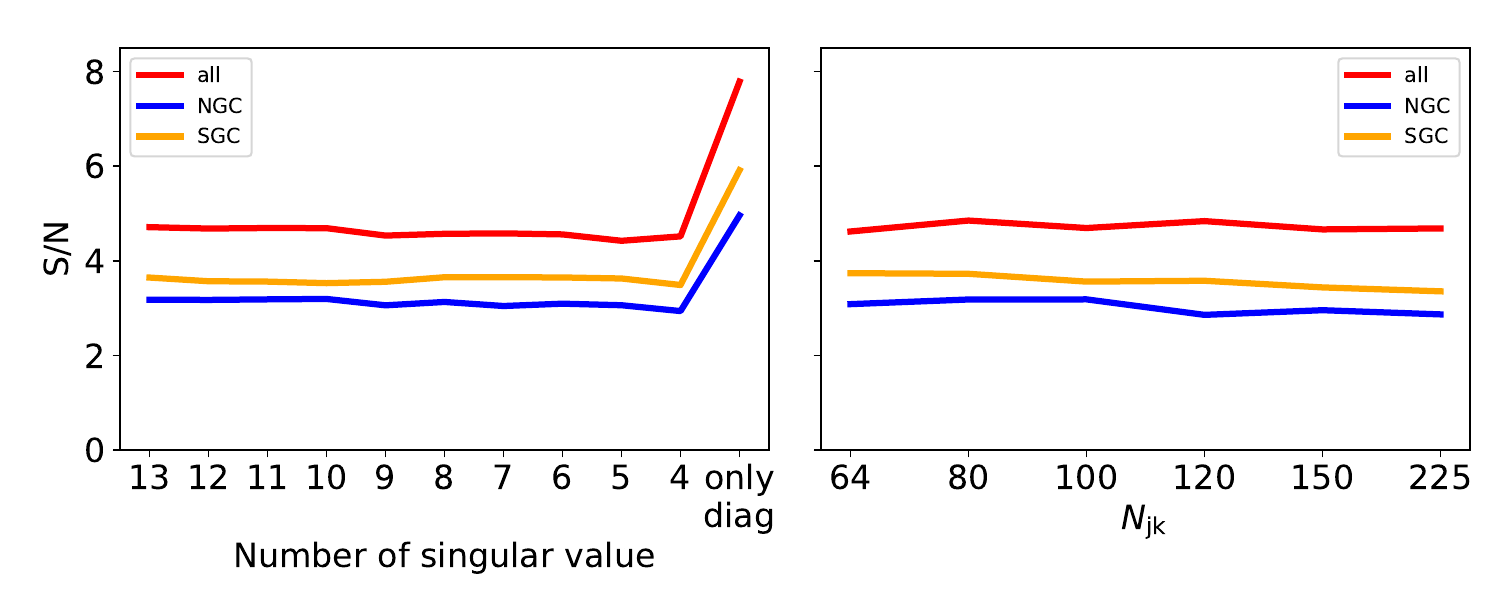}
	\caption{\textbf {Left}: The S/N as a function of the number of eigenmodes included in the SVD.  For comparison, we also show the S/N by using only the diagonal elements of the covariance (the last data points). If off-diagonal elements are neglected, the S/N will be overestimated. \textbf {Right}: The S/N as a function of $N_{\rm JK}$. \label{fig:app_SN_ev_njk}}
	\end{figure*} 
		
	Fig. \ref{fig:baseline_model_corva} shows that the covariance matrix (CM) has large cross-correlation at larger $r$. Therefore we  test in two ways whether the estimated CM  with 100 jackknife samples is sufficiently accurate. (1) We quantify the impact of CM on the S/N of the kSZ measurement, with two methods. 
The ratio of the maximum and minimum eigenvalues of the baseline CM (Fig. \ref{fig:baseline_model_corva}) is larger than $100$. So we first use  SVD to pseudo-inverse the matrix and estimate the S/N. The estimated S/N as a function of the number of eigenmodes are shown in Fig. \ref{fig:app_SN_ev_njk} (\textbf {Left}). They are consistent with each.  Therefore, the numerical method calculating the inverse matrix does not influence the final result. (2) We measure the S/N as a function of the number of jackknife samples (Fig. \ref{fig:app_SN_ev_njk} (\textbf {Right})). The estimation of S/N is stable around $N_{\rm jk}=100$. Therefore we conclude that our estimation of CM is sufficiently accurate for the data analysis and quantification of S/N.

\section{Choices of cluster sample} \label{app:choices_of_galaxy_sample}
	\begin{figure*}
	\centering
	\includegraphics[width=1\textwidth]{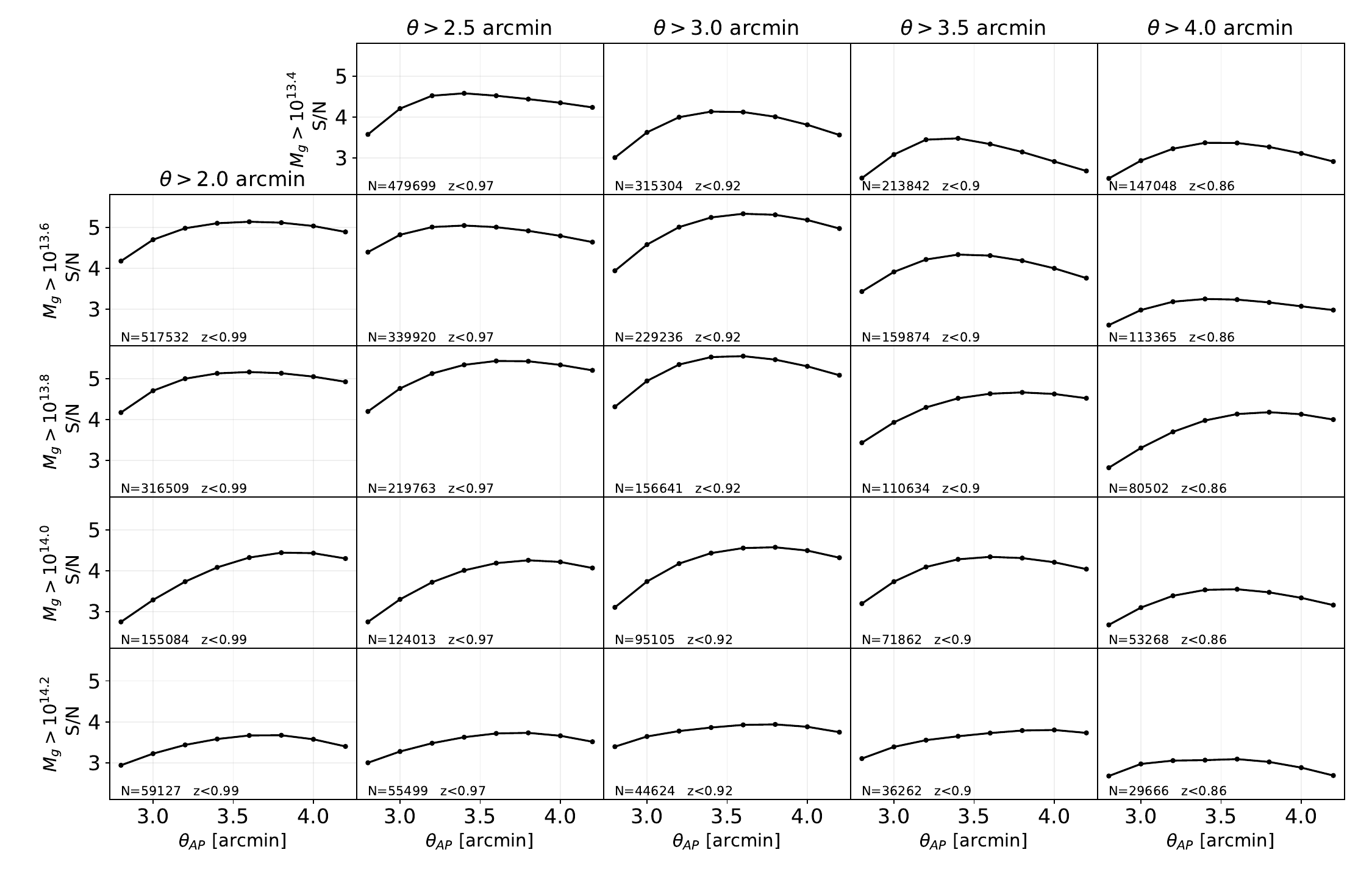}
	\caption{ The S/N of different clusters samples with varying mass and angular size threshold, as a function of the size of AP filter. The total number of clusters and the maximum redshift of each sample are also shown. \label{fig:APP_sample_sn}}
	\end{figure*}
The choice of galaxy sample influences the measured signal, its statistical significance, and residual systematic errors. There are many factors, such as the cluster mass, angular size and redshift errors, to consider in selecting clusters. Furthermore, the size (total number) of cluster sample matters.  Therefore, we need to keep a balance between these considerations. Fig. \ref{fig:APP_sample_sn} shows S/N of different samples with varying angular radius threshold and the mass threshold, as a function of the AP filter size.  When the size of the AP filter is comparable to the angular radius, S/N is maximized. This is expected, since too small AP filter size filters away too much kSZ signal, and too large AP filter size leaves too much primary CMB.  The highest S/N is $5.6$,  for the sample of clusters with angular radius larger than $3^{'}$, $M\geq 10^{13.8}M_{\odot}h^{-1}$, and AP filter size $3.6^{'}$. The measured pairwise kSZ signal of this sample is shown in Fig. \ref{fig:pksz_highest_sn}.

\section{Redshift dependence} \label{app:redshift}
Redshift evolution of the kSZ effect contains valuable information, on both the baryon content of clusters, and the structure growth rate. Therefore we attempt to divide the baseline cluster sample into two redshift bins and extract such redshift information. The first redshift bin has $z<0.31$, with $\langle M\rangle=10^{14.07} M_{\odot} h^{-1}$ and $\langle z\rangle= 0.215$. The second has  $0.31<z< 0.88$, with $\langle M\rangle=10^{14.27} M_{\odot} h^{-1}$ and $\langle z\rangle= 0.427$.  The two bins have equal number of clusters. 
\begin{figure}
	\centering
	\includegraphics[width=0.45\textwidth]{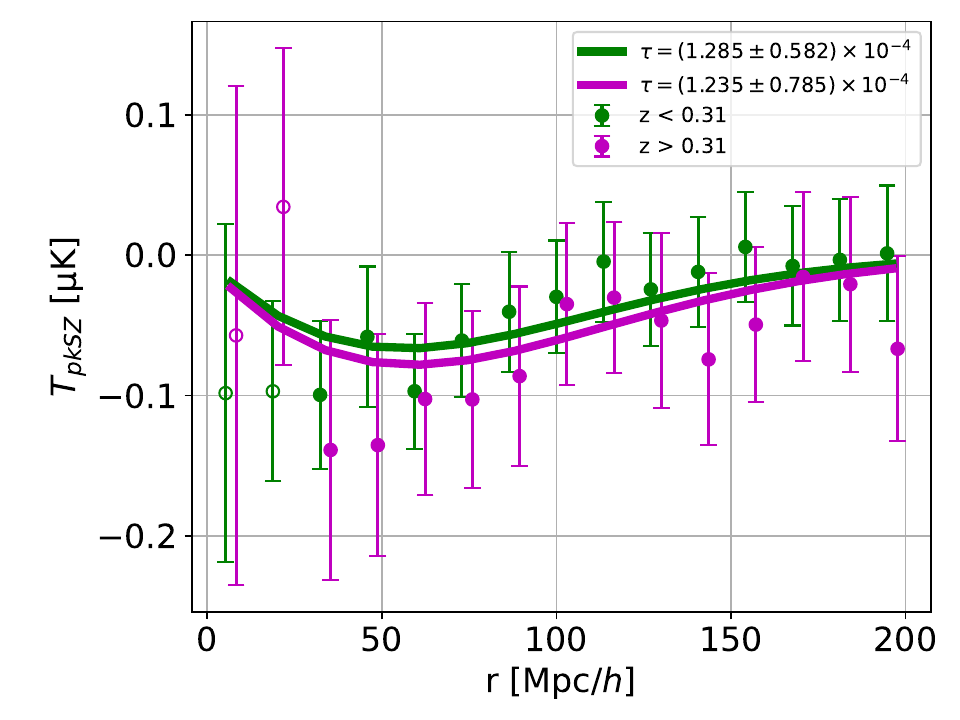}
	\caption{The pairwise kSZ measurements of clusters at two reshift bins. \label{fig:results_z_dependence}}
    	\end{figure}
The pairwise kSZ measurements are shown in Fig. \ref{fig:results_z_dependence}. We constrain $\bar{\tau}=(1.29 \pm 0.58) \times 10^{-4} (2.2 \sigma)$ for the low redshift bin and $\tau=(1.24 \pm 0.78) \times 10^{-4} (1.6 \sigma)$ for the high redshift bin. We detect no significant evolution in the mean optical depth. Since the detection significance is low, we are not able to robustly quantify the redshift evolution or correct the difference in their mass and other redshift related selection effects. This issue shall be investigated robustly with future higher-resolution and higher-sensitivity CMB experiments, and cluster samples with spectroscopic redshifts.

\end{document}